%% file: main.tex
\RequirePackage[columnwise]{lineno}
\documentclass[prd,aps,twocolumn,superscriptaddress,showpacs,showkeys,amsmath,amssymb,floatfix]{revtex4}
\usepackage[colorlinks=true,citecolor=blue,filecolor=blue,linkcolor=blue,urlcolor=blue,pdftex]{hyperref}

\usepackage{xspace} 
\usepackage[usenames,dvipsnames]{color}
\usepackage{booktabs,graphicx,mathrsfs,verbatim,amsmath,units,soul}
\usepackage{xspace} 
\usepackage{upgreek}
\usepackage{gensymb}
\usepackage{amsmath}
\usepackage{tabularx}
\usepackage[none]{hyphenat}
\usepackage{multirow}
\usepackage{layouts}
\usepackage{printlen}

\newcommand{\ton}{t\xspace}

\newcommand{\zCoord}{$z$\xspace}

\newcommand{\secref}{Sec.\xspace}
\newcommand{\figref}{Fig.\xspace}

\input{affiliations.tex}

\begin{document}

\title{XENON1T Dark Matter Data Analysis: \\ Signal Reconstruction, Calibration and Event Selection}


\input{authors.tex}

\date{\today}

\begin{abstract}

The XENON1T experiment at the Laboratori Nazionali del Gran Sasso is the most sensitive direct detection experiment for dark matter in the form of weakly interacting particles (WIMPs) with masses above $6\,$GeV/$c^2$ scattering off nuclei. The detector employs a dual-phase time projection chamber with 2.0 metric tons of liquid xenon in the target. A one metric $\mathrm{ton}\times\mathrm{year}$ exposure of science data was collected between October 2016 and February 2018. This article reports on the performance of the detector during this period and describes details of the data analysis that led to the most stringent exclusion limits on various WIMP-nucleon interaction models to date. In particular, signal reconstruction, event selection and calibration of the detector response to nuclear and electronic recoils in XENON1T are discussed.
\end{abstract}


\maketitle

\input{Introduction.tex}

\input{StabilityDataTaking.tex}

\input{Acquisisition_Reconstruction.tex}

\input{SignalCorrections.tex}

\input{Cuts.tex}

\input{Discrimination.tex}

\input{Outlook.tex}

\acknowledgments
We gratefully acknowledge support from the National Science Foundation, Swiss National Science Foundation, German Ministry for Education and Research, Max Planck Gesellschaft, Deutsche Forschungsgemeinschaft, Netherlands Organisation for Scientic Research (NWO), Netherlands eScience Center (NLeSC) with the support of the SURF Cooperative,Weizmann Institute of Science, Israeli Centers Of Research Excellence (I-CORE), Pazy-Vatat, Initial Training Network Invisibles (Marie Curie Actions, PITNGA-2011-289442), Fundacao para a Ciencia e a Tecnologia, Region des Pays de la Loire, Knut and AliceWallenberg Foundation, Kavli Foundation, and Istituto Nazionale di Fisica Nucleare. Data processing is performed using infrastructures from the Open Science Grid and European Grid Initiative. We are grateful to Laboratori Nazionali del Gran Sasso for hosting and supporting the XENON project.

\bibliographystyle{apsrev}
\bibliography{bibliography.bib}

\end{document}

%% file: affiliations.tex
\newcommand{\bologna}{\affiliation{Department of Physics and Astronomy, University of Bologna and INFN-Bologna, 40126 Bologna, Italy}}

\newcommand{\chicago}{\affiliation{Department of Physics \& Kavli Institute for Cosmological Physics, University of Chicago, Chicago, IL 60637, USA}}

\newcommand{\coimbra}{\affiliation{LIBPhys, Department of Physics, University of Coimbra, 3004-516 Coimbra, Portugal}}

\newcommand{\columbia}{\affiliation{Physics Department, Columbia University, New York, NY 10027, USA}}

\newcommand{\lngs}{\affiliation{INFN-Laboratori Nazionali del Gran Sasso and Gran Sasso Science Institute, 67100 L'Aquila, Italy}}

\newcommand{\mainz}{\affiliation{Institut f\"ur Physik \& Exzellenzcluster PRISMA, Johannes Gutenberg-Universit\"at Mainz, 55099 Mainz, Germany}}

\newcommand{\heidelberg}{\affiliation{Max-Planck-Institut f\"ur Kernphysik, 69117 Heidelberg, Germany}}

\newcommand{\munster}{\affiliation{Institut f\"ur Kernphysik, Westf\"alische Wilhelms-Universit\"at M\"unster, 48149 M\"unster, Germany}}

\newcommand{\nikhef}{\affiliation{Nikhef and the University of Amsterdam, Science Park, 1098XG Amsterdam, Netherlands}}

\newcommand{\nyuad}{\affiliation{New York University Abu Dhabi, 129188 Abu Dhabi, United Arab Emirates}}

\newcommand{\purdue}{\affiliation{Department of Physics and Astronomy, Purdue University, West Lafayette, IN 47907, USA}}

\newcommand{\rpi}{\affiliation{Department of Physics, Applied Physics and Astronomy, Rensselaer Polytechnic Institute, Troy, NY 12180, USA}}

\newcommand{\rice}{\affiliation{Department of Physics and Astronomy, Rice University, Houston, TX 77005, USA}}

\newcommand{\stockholm}{\affiliation{Oskar Klein Centre, Department of Physics, Stockholm University, AlbaNova, Stockholm SE-10691, Sweden}}

\newcommand{\subatech}{\affiliation{SUBATECH, IMT Atlantique, CNRS/IN2P3, Universit\'e de Nantes, Nantes 44307, France}}

\newcommand{\torino}{\affiliation{INFN-Torino and Osservatorio Astrofisico di Torino, 10125 Torino, Italy}}

\newcommand{\ucla}{\affiliation{Physics \& Astronomy Department, University of California, Los Angeles, CA 90095, USA}}

\newcommand{\ucsd}{\affiliation{Department of Physics, University of California, San Diego, CA 92093, USA}}

\newcommand{\wis}{\affiliation{Department of Particle Physics and Astrophysics, Weizmann Institute of Science, Rehovot 7610001, Israel}}

\newcommand{\zurich}{\affiliation{Physik-Institut, University of Zurich, 8057  Zurich, Switzerland}}

\newcommand{\paris}{\affiliation{LPNHE, Universit\'{e} Pierre et Marie Curie, Universit\'{e} Paris Diderot, CNRS/IN2P3, Paris 75252, France}}

\newcommand{\freiburg}{\affiliation{Physikalisches Institut, Universit\"at Freiburg, 79104 Freiburg, Germany}}

\newcommand{\lal}{\affiliation{LAL, Universit\'e Paris-Sud, CNRS/IN2P3, Universit\'e Paris-Saclay, F-91405 Orsay, France}}

\newcommand{\naples}{\affiliation{Department of Physics ``Ettore Pancini'', University of Napoli and INFN-Napoli, 80126 Napoli, Italy}}

\newcommand{\nagoya}{\affiliation{Kobayashi-Maskawa Institute for the Origin of Particles and the Universe, Nagoya University, Furo-cho, Chikusa-ku, Nagoya, Aichi 464-8602, Japan}}

%% file: authors.tex

\author{E.~Aprile}\columbia
\author{J.~Aalbers}\stockholm\nikhef
\author{F.~Agostini}\bologna
\author{M.~Alfonsi}\mainz
\author{L.~Althueser}\munster
\author{F.~D.~Amaro}\coimbra
\author{V.~C.~Antochi}\stockholm
\author{F.~Arneodo}\nyuad
\author{L.~Baudis}\zurich
\author{B.~Bauermeister}\stockholm
% ...
\collaboration{XENON Collaboration}
\email[]{xenon@lngs.infn.it}
\noaffiliation

%% file: Introduction.tex
\section{Introduction}
\label{sec:Introduction}


The existence of a non-luminous, massive matter component beyond the standard model, called dark matter, is evidenced by numerous astrophysical observations \cite{Bertone:2016nfn}. Among the best motivated dark matter candidates are weakly interacting massive particles (WIMPs)~\cite{Roszkowski_2018, JUNGMAN1996195}. Ultra-sensitive Earth-based detectors provide one possible approach to the direct detection of WIMPs as the particles are expected to scatter off the detector's target nuclei~\cite{Undagoitia:2015gya,Baudis_2016}. This induces nuclear recoils with mean energies in the order of a few keV.

Experiments that employ the liquid xenon (LXe) time projection chamber (TPC) technology are leading the search for elastic WIMP-nucleon interactions for masses from a few GeV/$c^2$ up to the TeV/$c^2$ scale \cite{xe1t_comb, LUX_SI_2017, PandaX_SI_2017}.
The XENON1T experiment \cite{xe1t_instru} has placed the most stringent upper limit on the spin-independent WIMP-nucleon cross section for masses above 6$\,$GeV/$c^2$ with a minimum of $4.1\times10^{-47}\,$cm$^2$ at 30$\,$GeV/$c^2$ and 90\% confidence level~\cite{xe1t_comb}.

Together with \cite{xe1t_long_analysis_2} this document reports on  the analysis methods employed for the spin-independent dark matter search with XENON1T and subsequent results \cite{xe1t_pion, xe1t_SD}. While the present article describes the techniques of signal reconstruction, event selection and detector calibration, details on the detector response model, the WIMP signal and background models, and the statistical inference are presented in \cite{xe1t_long_analysis_2}.


The XENON1T detector was hosted by the Laboratori Nazionali del Gran Sasso (LNGS) and was operated with a total of $\sim3200\,$kg of ultra-pure LXe, with $(2004\pm5)\,$kg contained in the TPC as active target
and the remainder used for shielding.
Additional shielding from ambient radioactivity was provided by a $10\,$m tall and $9.6\,$m diameter water tank that served as active 
Cherenkov muon veto \cite{xe1t_muonveto} and passive shielding. The TPC was cylindrically shaped with a diameter of $96\,$cm and height of $97\,$cm. The top and bottom surfaces were instrumented by arrays of 248~Hamamatsu  R11410-21 low-background photomultiplier tubes (PMTs) of 3" diameter in total~\cite{PMTs_xe1t, PMTs_screen}.

Particles scatter off xenon atoms inducing either nuclear recoils (NR) or electronic recoils (ER).
The recoil energy is measured by detecting signals from excitation and ionization of xenon atoms, where the relative contribution of these two channels depends on the recoil type.
Excited xenon atoms emit $178\,$nm scintillation light~\cite{basov_1979} which is observed by the PMTs and is referred to as \textit{S1} signal. The ionization electrons are extracted from the interaction site by an electric field and are drifted towards the 
liquid-gas interface at the top of the TPC where they are extracted into the gas phase by another, stronger electric field and create the proportional scintillation \textit{S2} signal~\cite{PropScintillation}.
S1 and S2 signals are anti-correlated due to recombination of electrons with ions creating excited xenon atoms. The two signals are temporally separated by the drift time of the electrons, from which the $z$-coordinate of the interaction is reconstructed. The $x$ and $y$ coordinates are inferred from the light pattern of the S2 signal on the top PMT array. The S2/S1-ratio is used to discriminate between ER background ($\gamma$ and $\beta$ radiation) and signal-like NR events (WIMPs and neutrons).

XENON1T was operated stably for more than one year from October 2016 to February 2018 (Sec.\,\ref{sec:data_stab}). The raw data acquired during this period were converted into physical quantities such as peak amplitude, area, width etc. by means of a data processor which is described in Sec.\,\ref{sec:paxfax}, together with an event simulation framework used to evaluate the processor's performance. The processor also reconstructs the interaction position of each event (Sec.\,\ref{sec:ProRec}) and applies corrections to the measured signals to account for spatial dependencies (Sec.\,\ref{sec:Sign_Corr}). The search for WIMP dark matter is based on the selection of a clean sample of single-scatter events (Sec.\,\ref{sec:Cuts}) inside a central fiducial volume which features a reduced background level (Sec.\,\ref{sec:cuts:fiducial}). Finally, we present the detector response to NR and ER events in Sec.\,\ref{sec:Discrimination}.


%% file: StabilityDataTaking.tex
\section{Detector Operation and Stability}
\label{sec:data_stab}

\subsection{Science Runs}
\label{subsec:datataking}

\begin{figure*}
\centering
\includegraphics[width=1\textwidth]{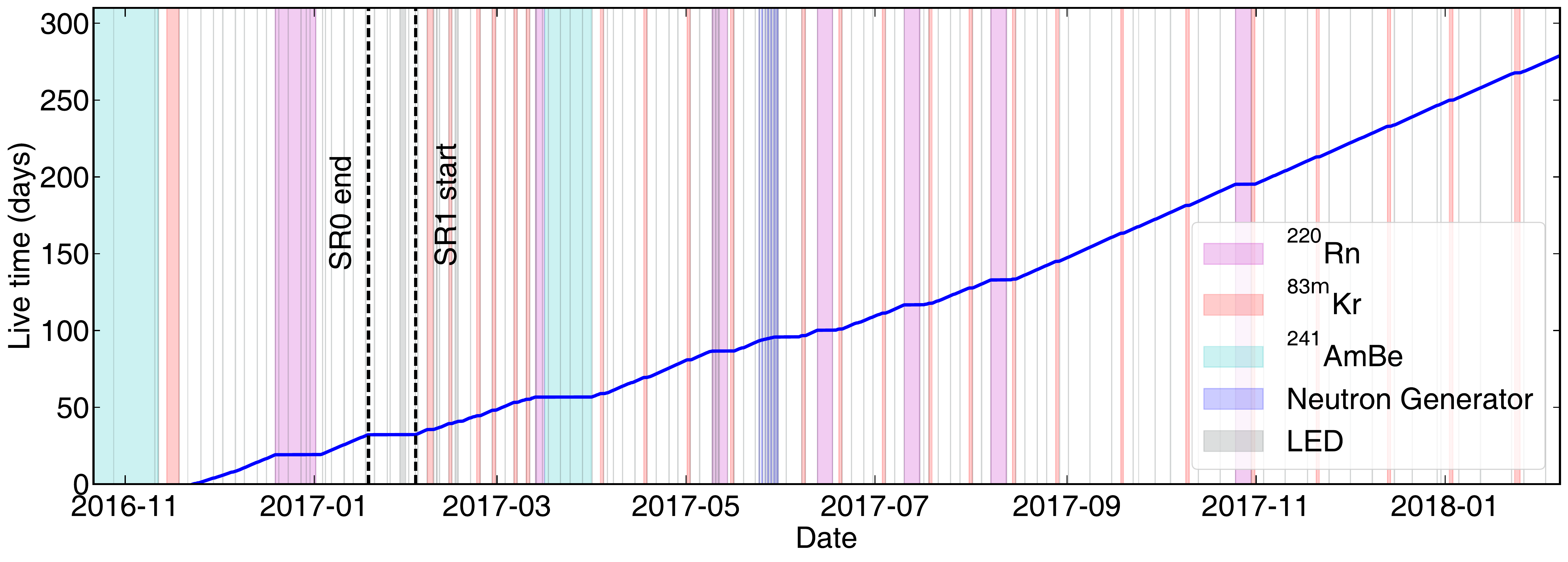}
\caption{Accumulated data live time acquired with the XENON1T detector in dark matter search mode and corrected for data quality conditions (Sec.\,\ref{sec:cuts:LivetimeReduction}). The black dashed vertical lines mark the end of SR0 and the start of SR1. The various colored bands represent periods of detector calibration with $^{220}$Rn (magenta), $^{83\mathrm{m}}$Kr (red), $^{241}$AmBe (cyan), neutron generator (blue) sources and with blue LED light (grey).}
\label{fig:datataking}
\end{figure*}

\begin{table*}
    \centering
    \begin{tabular}{|c|c|c|}
     \hline\textbf{Data live time attributes} & \textbf{SR0} & \textbf{SR1} \\\hline
    Science data live time [days] & 32.1 & 246.7 \\
    Science data live time reduction due to data quality [\%] & 13.6 & 6.9\\
    Calibration live time  [days] & 39.2 & 83.6 \\\hline
    \textbf{Detector operation parameters} & \textbf{SR0} & \textbf{SR1}\\\hline
    Drift field: FV-averaged (mean $\pm$ $1\sigma$) [V/cm] & $120\pm8$ & $81\pm6$\\
    Average electron lifetime [$\mu$s] & 290 & 641\\
    Extraction field: FV-averaged (mean $\pm$ $1\sigma$) [kV/cm] & $8.1\pm0.1$ & $8.1\pm0.1$ \\
    Number (Fraction [\%]) of excluded PMTs & 35 (14.1) & 36 (14.5) \\
    Liquid xenon temperature (mean $\pm1\sigma$) [$^\circ$C] & $-96.07\pm0.04$ & $-96.02\pm0.02$\\
    Xenon gas pressure (mean $\pm1\sigma$) [bar] & $1.934\pm0.001$ & $1.938 \pm0.001$\\
    Charge yield: max. deviation from mean [\%] & 2 & 2\\
    Light yield: max. deviation from mean [\%] & 1 & 1\\
    Single electron (SE) gain (mean $\pm1\sigma$) [PE/SE] & $27.2\pm0.9$ & $28.2\pm0.1$\\
    \hline
    \end{tabular}
    \caption{Comparison of data live time attributes and detector operation parameters among the two science runs of XENON1T. The science data live time reduction refers to the fraction of data removed due to data quality criteria (Sec.\,\ref{sec:cuts:LivetimeReduction}). The mean and $1\sigma$ values of the detector operation parameters correspond to their temporal distribution, except for the electric fields where the mean and $1\sigma$ refer to the spatial homogeneity. Note that PE is the abbreviation of photoelectron.}
    \label{tab:SR_conditions}
\end{table*}


XENON1T performed its first science run (SR0) between October 2016 and January 18th, 2017 when a magnitude 5.7 earthquake temporarily interrupted the detector's operation. In SR0, the drift field yielded $120\,$V/cm with a RMS of $8\,$V/cm averaged over a $1.3\,$\ton fiducial volume (FV) (see Sec.\,\ref{sec:cuts:fiducial}). The second science run (SR1) was launched February 2nd, 2017. SR1 featured a lower drift field of $81\,$V/cm with a RMS of $6\,$V/cm. The field was reduced as a consequence of high voltage instabilities occurring after the earthquake. Compared to the improvement of the electron lifetime in SR1 (Sec.\,\ref{sec:ELifetimeCorr}), this field reduction has negligible impact on the analysis results in view of energy resolution and signal acceptance. The drift field is taken into account in the signal and background models \cite{xe1t_long_analysis_2} and the electric field values are determined from simulations, using finite element (COMSOL Multiphysics \cite{Comsol}) and boundary element methods (KEMField \cite{Kemfield_sim}).
SR1 continued until February 24th, 2018, marking over one year of stable data-taking, as shown in Fig.\,\ref{fig:datataking}. The blue line presents the accumulated dark matter live time versus calendar time, corrected for the data quality conditions described in Sec.\,\ref{sec:cuts:LivetimeReduction} which reduce the live time by 13.6\% and 6.9\% in SR0 and SR1, respectively (see Tab.\,\ref{tab:SR_conditions}).

In addition to science data acquisition, various calibration campaigns were performed as shown in Fig.\,\ref{fig:datataking} by the vertical colored bands. The $^{220}$Rn decay chain includes the $^{212}$Pb $\beta$-decay, which is employed to calibrate the detector response to low energy ER events for background modeling and to derive event selection criteria~\cite{xe100_rn220}. The metastable isotope $^{83\mathrm{m}}$Kr decays into its ground state by emitting $32.1\,$ and $9.4\,$keV conversion electrons with half lives of $1.83\,$h and $157\,$ns~\cite{bnl_nndc}, respectively. Those decays are used to monitor spatial and time dependencies of detector signals (Sec.\,\ref{sec:Sign_Corr}) and reconstruct signal positions (Sec.\,\ref{pos:cor}).
The $^{220}$Rn and $^{83\mathrm{m}}$Kr sources are injected into XENON1T via the gas purification loop. Shortly after injection, the isotopes are distributed homogeneously throughout the TPC~\cite{lux_kr83m}. To avoid potential impact on the position and signal correction from the non-uniformity of $^{83\mathrm{m}}$Kr events at the beginning of injection, the first hour data was not used for correction map generation. The acquisition of dark matter data was resumed after the calibration campaigns when the trigger rate had fallen to background level. In this science data we also observed a trace amount of $^{83\mathrm{m}}$Kr events which presumably was caused by a malfunction of the source valve. These events are not in the energy region of interest for standard dark matter searches but can be used to monitor the detector response throughout SR1.

\begin{figure*}
\centering
\includegraphics[width=\textwidth]{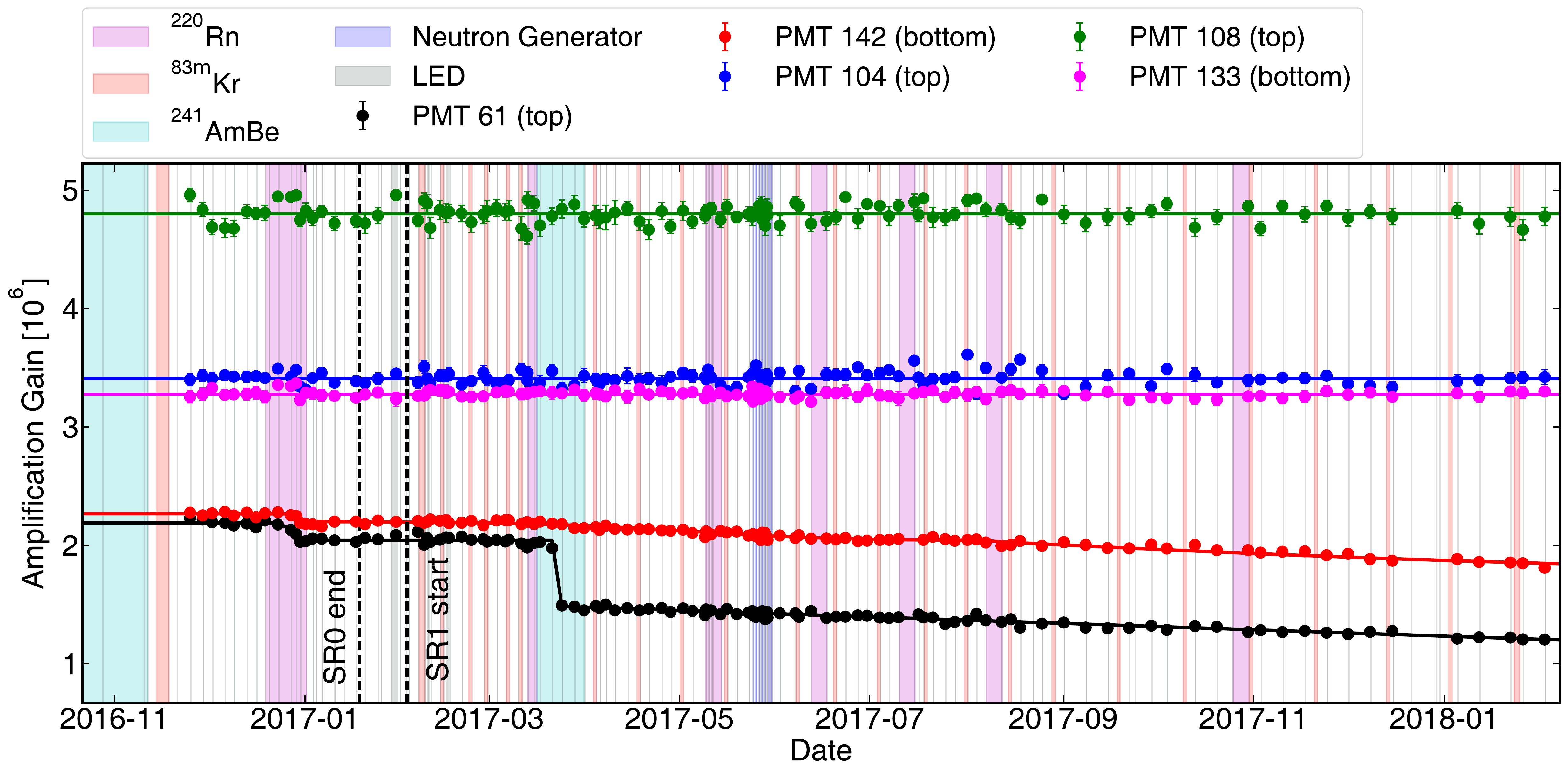}
\caption{PMT gains measured by LED calibrations as a function of time for three representative stable PMTs (green, blue and magenta) and two examples where the gain decreased due to small vacuum leaks (red and black).}
\label{fig:pmtgains}
\end{figure*}

A deuterium-deuterium plasma fusion neutron generator~\cite{ng_rafael_2017} and a vessel containing a $^{241}$AmBe~\cite{xe1t_instru} source were immersed into the water tank next to the cryostat to calibrate the detector response to NR events for signal modeling.

Additionally, monoenergetic $\gamma$-lines from radio-impurities in the detector materials ($^{60}$Co, $^{40}$K) and xenon isotopes excited after neutron calibrations ($^{129\mathrm{m}}$Xe, $^{131\mathrm{m}}$Xe), as well as the combined $41.5\,$keV conversion electron signal from $^{83\mathrm{m}}$Kr were used to monitor the detector stability and calibrate the energy scale over three orders of magnitude \cite{xe1t_instru}.

%


\subsection{Detector Stability}
\label{subsec:pmts}

During both science runs, the detector was operated under stable thermodynamic conditions that guarantee constant signal sizes. 

In SR0 (SR1) the $1\sigma$ fluctuations of the xenon gas pressure were within 0.05\% (0.05\%). The pressure were uniformly distributed inside detector which also reflected on the stability of the single electron (SE) gain, i.e. the size of S2 signals from SE extracted from the liquid into the gas phase. The SE gain is determined by the gas pressure, the LXe level and the electric extraction field. SE signals are identified similarly to the procedure described in \cite{xe100_se} and the extracted SE gain has an average of $27.2\,$photoelectrons(PE)/SE ($28.2\,$PE/SE) and a $1\sigma$ time variation of 3.2\% (0.5\%) in SR0 (SR1) and a SE $1\sigma$ width of $7.3\,$PE/SE ($7.4\,$PE/SE) with a time stability of 11\% (5\%). The slightly larger variation in SR0 is caused by small fluctuations of the LXe level. 
The fluctuation of the LXe temperature were within 0.04\% (0.02\%) for SR0 (SR1). The temperature difference between the top and bottom of the TPC was within 0.25$^\circ$C. The potential impact of the temperature variation on signal yields was taken into account by the signal corrections in Sec.\,\ref{sec:Sign_Corr}.
Tab.\,\ref{tab:SR_conditions} lists the most important detector operation parameters and their $1\sigma$ spread around the temporal mean value.

The PMT gains were measured weekly with a pulsed blue light-emitting diode (LED) configured to stimulate the emission of a low number of PEs from the photocathode and the gains are extracted using the model independent approach described in \cite{gainmodel}. Fig.\,\ref{fig:pmtgains} shows the gain evolution of 3 stable PMTs (104, 108, 133) that are representative for the majority of PMTs in the XENON1T TPC.
In SR0 (SR1), 35 (36) PMTs were excluded from analysis~\cite{xe1t_first}, with 15 (15) in the top and 20 (21) in the bottom array. Nearly all of these PMTs suffered from vacuum leaks causing decreased performance such as light emission and afterpulses, requiring the bias voltage of the PMT to be lowered, eventually to zero. The criterion for exclusion from analysis is a single photoelectron (SPE) acceptance smaller than 50\%~\cite{xe1t_instru}. Although vacuum leaks lead to the decrease of PMT gains, 19 PMTs with small leaks were operated successfully throughout the science runs and 
their gains were monitored and corrected.
Two examples, PMTs 61 and 142, are shown in Fig.\,\ref{fig:pmtgains}. The gain evolution is modeled empirically by Fermi-Dirac functions that take into account the time when the PMT high voltage (HV) was lowered. The standard deviation of the measured gain with respect to the model is within a few percent for both stable and decreasing PMT gains and is dominated by statistical uncertainties. Compared to the approximately 27\% resolution of the PMT response to SPEs~\cite{PMTs_xe1t}, the systematic uncertainty of the gain has negligible impact on the energy resolution.

The temporal stability of the S1 and S2 signals is further confirmed by monitoring the light and charge yield (LY and CY) evolution over time. Using data from mono-energetic sources between $9.4\,$keV ($^{83\mathrm{m}}$Kr) and $5.6\,$MeV ($^{222}$Rn), the measured S1 and S2 signals per incident energy are evaluated following the same procedure as in \cite{xe1t_instru}. The values are stable throughout both science runs with maximum deviations from the mean of 1\% and 2\%, respectively.


%% file: Acquisisition_Reconstruction.tex
\section{Signal Reconstruction and Simulation}
\label{sec:paxfax}


PMT signals exceeding a channel-specific threshold above the baseline, accepting on average 93\% of SPE signals, are digitized at a rate of $10^8\,$samples/second by the data acquisition (DAQ) system \cite{xe1t_daq}. These signals are referred to as \textit{pulses}. An online event builder groups pulses into events using a simplified algorithm to trigger on S1 and S2 candidates and stores a $1\,$ms window around each trigger. During offline processing by the custom developed data processor \textit{PAX}~\cite{pax}, pulses are further segmented into smaller intervals, denoted as \textit{hits}, by separating individual signals, which may have been grouped into the same pulse waveform. Hits from different PMT channels are grouped into clusters in time, referred to as \textit{peaks}, and corresponding to individual ionization or scintillation signals. Properties of each peak, such as area, width, and height are computed by the processor. A peak is classified as S1 (S2) if its waveform rises sufficiently fast (slow) and has at least 3 (4) contributing PMTs. For S1s, only hits with maxima within a $100\,$ns window centered on the maximum of the sum-waveform for all channels are counted for the latter requirement.
Finally, each event is searched for a valid S1-S2 pairing, starting with the largest peaks of each type. These pairings are called \textit{interactions}.
A further reduction of the processor output is performed by the software package \textit{HAX}~\cite{hax}. Signal corrections (Sec.\,\ref{sec:Sign_Corr}) as well as other higher-level algorithms are included at this level.

\begin{figure}
    \centering
    \includegraphics[width=1\linewidth]{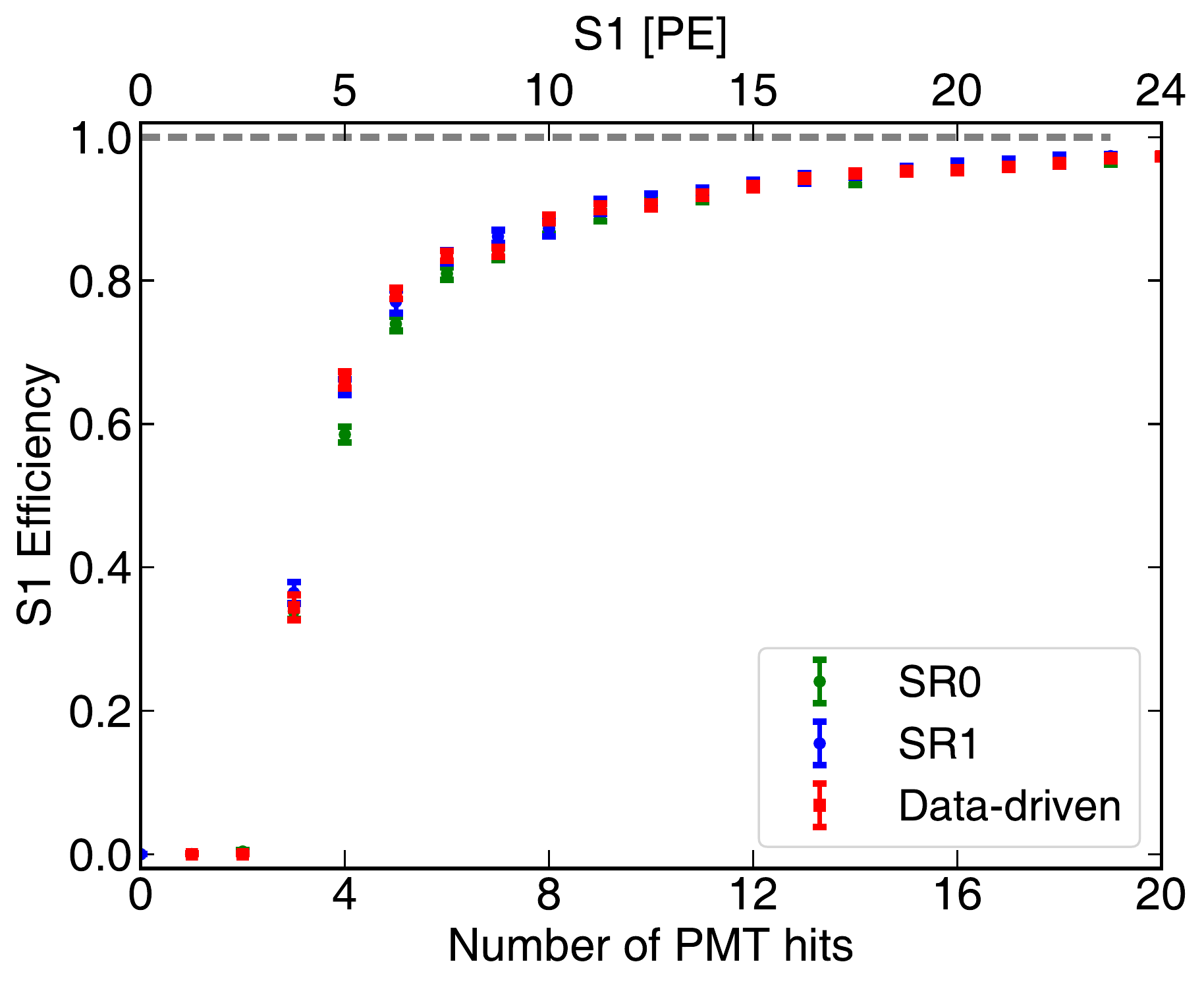}
    \caption{S1 signal reconstruction efficiency estimated from waveform simulations, as a function of number of PMT hits. Hits are converted into S1 signal size (top axis) using a double PE emission probability at the photocathode of 21.9\%~\cite{xe1t_long_analysis_2}.
    The data-driven efficiency from $^{220}$Rn calibration is overlaid for comparison.}
    \label{fig:S1Acceptance}
\end{figure}

The performance of the data processor is studied using emulated PMT signals from a waveform simulator. The simulation employs data-driven models of XENON1T detector specific properties like the scintillation light pulse shape, the spatial dependence of the light collection efficiency, the diffusion of electrons during drift, the time profile of SE, PMT afterpulses, the SEs generated by photo-ionization of impurities, and the electronic noise. The simulated data is validated by comparison to $^{83\mathrm{m}}$Kr and neutron calibration data and provides the means to optimize
the reconstruction algorithms in the data processor and quantify their performance.

The S1 signal reconstruction efficiency is determined from simulated waveforms and is shown in Fig.\,\ref{fig:S1Acceptance}. The efficiency is a function of the number of PMT hits, which at low energies is equivalent to the number of detected photons. The conversion into the S1 peak area (shown in the top axis of Fig.\,\ref{fig:S1Acceptance}) assumes a double electron emission probability at the PMT's photocathode of 21.9\% \cite{xe1t_long_analysis_2}. The efficiency's uncertainty is estimated from the simulation by varying the data-driven model parameters S1 width, PMT afterpulse rates and rate of photo-ionization at the gate within their uncertainties. 
The results are cross-checked with a data-driven method, 
of which subsets of hits from a large S1 are selected to build an artificial low energy S1 and the efficiency is calculated based on these low energy artificial S1s.
The result from the data-driven method using $^{220}$Rn calibration is in agreement with simulations as shown in Fig.\,\ref{fig:S1Acceptance}. 
Compared to SR1, the S1 efficiency for SR0 is slightly smaller at the threshold due to a higher PMT noise level in the first third of SR0 dark matter data acquisition that could be reduced by installing low-pass filter boxes at the PMT high voltage modules \cite{xe1t_daq}.

The S2 trigger efficiency of the offline event builder is determined by applying the trigger algorithm to simulated S2 waveforms that were generated homogeneously throughout the TPC.
The analysis threshold for S2 signal sizes is fixed to $200\,$PE where the trigger efficiency yields $(99.8^{+0.2}_{-0.6})$\% and $(99.4^{+0.4}_{-0.7})$\% in SR0 and SR1, respectively, and the S2 size is defined using the sum of the signals from the top and bottom PMT arrays. Recently, a data-driven method became available and resulted in about 4\% smaller efficiencies at the S2 signal threshold~\cite{xe1t_daq}. However, this is expected to have no noticeable impact on the dark matter search since the S1 signal reconstruction efficiency is the dominating parameter determining the detection threshold.

\begin{figure*}
    \centering
    \includegraphics[width=0.95\textwidth]{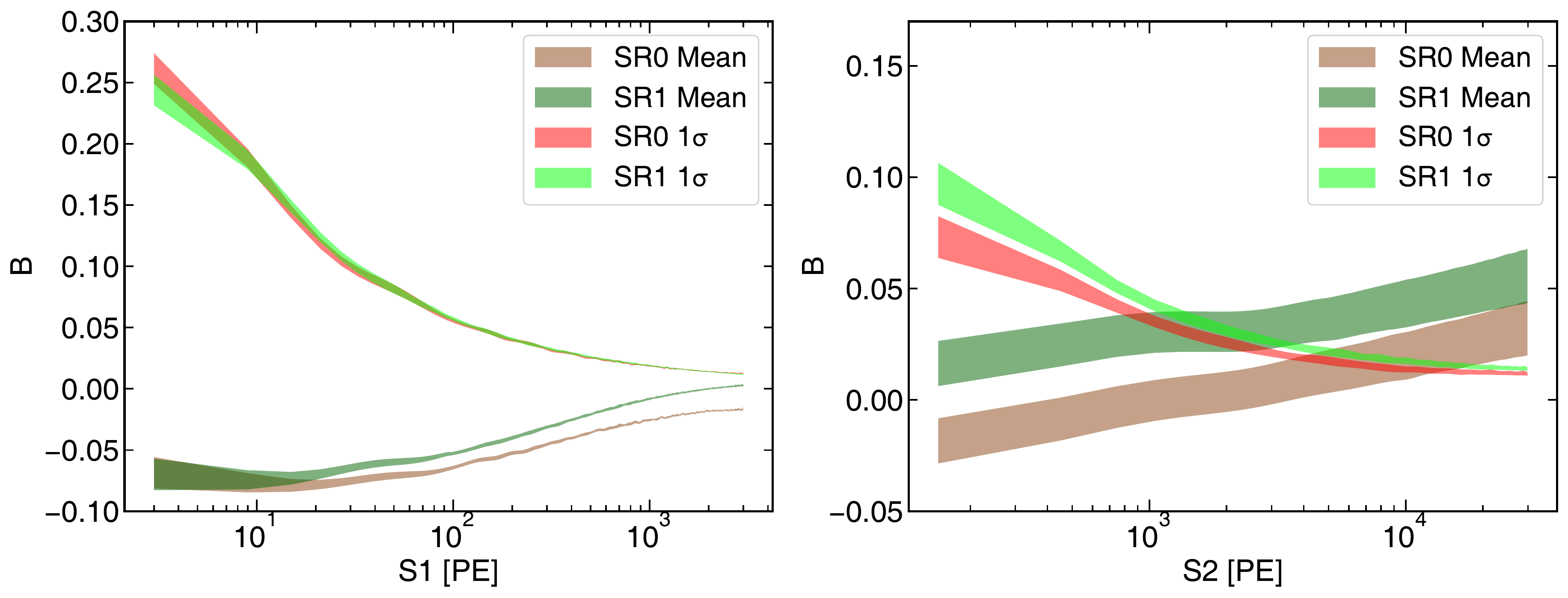}
    \caption{S1 (left) and S2 (right) signal size reconstruction bias $B$ and its $1\sigma$ width as estimated from simulations. The bands are derived by varying the data-driven model parameters within their uncertainties and hence represent the credible region of the shown values.}
    \label{fig:recon_bias}
\end{figure*}

The bias $B$ in reconstructing peak area is estimated by simulating homogeneously distributed S1 and S2 peaks. The expected number $A_\mathrm{true}$ of detected PEs of a given peak is compared with the number $A_\mathrm{rec}$ of reconstructed PEs:
\begin{equation} 
\label{eq_bias}
B = (A_\mathrm{rec} - A_\mathrm{true})/A_\mathrm{true}.
\end{equation}
The reconstruction bias is pre-dominantly caused by signals from photo-ionization, particularly coming from the gate electrode, and afterpulse signals in the PMTs. These signals can be merged to or cut-off from the primary peak.
Fig.\,\ref{fig:recon_bias} shows the mean and $1\sigma$ width
of the gaussian shaped distribution of $B$ as functions of S1 and S2 signal sizes. The uncertainty bands are estimated following the same procedure as for the S1 reconstruction efficiency. The mean bias and its width are very similar for the two science runs in the case of S1 signals. However, the two parameters are slightly higher for S2 signals measured in SR1 compared to those in SR0. This difference is most likely caused by a higher PMT afterpulse rate in SR1 as a result of increasing PMT vacuum leaks as explained in Sec.\,\ref{subsec:pmts}.

The S1 reconstruction efficiency and S1 and S2 signal reconstruction biases are input parameters for the signal and background response models~\cite{xe1t_long_analysis_2}. 

\section{Position reconstruction and related corrections}
\label{sec:ProRec}

Three-dimensional position reconstruction is one of the main advantages of dual-phase TPCs. 
Most radiogenic background events are located near the boundaries of the TPC and are rejected by selecting a radiopure inner fiducial volume (Sec.\,\ref{sec:cuts:fiducial}). In addition, accurate position reconstruction is required for the development of background models~\cite{xe1t_long_analysis_2} and for position-dependent signal corrections (Sec.\,\ref{sec:Sign_Corr}). 

\subsection{Position Reconstruction Methods}
\label{subsec:PosRec}

The vertical coordinate $z_\mathrm{obs}$ (the subscript obs indicating the position before correction as described in Sec.\,\ref{pos:cor}) of an interaction is determined by the electron drift velocity and the time difference between the prompt S1 and the delayed S2 signal. The origin of the coordinate is at the gate electrode and the TPC height extends down to $-97\,$cm.
Due to the diffusion of the electron cloud during the drift, multiple scatter events with close proximity in \zCoord are more difficult to separate at the bottom of the TPC than at the top. By identifying multiple scatter events in NR calibration data, the distance in \zCoord for the two interactions is determined. The distribution features a roll-off for small \zCoord distances varying with the interaction depth. The minimal value for which two scatters are separable with an acceptance of 50\% is found to increase from $2\,$mm at the gate to $7\,$mm at the cathode for S2 signals in the region of interest for dark matter searches ($\mathrm{S2}<25000\,$PE).

The horizontal position ([$x_\mathrm{obs}$, $y_\mathrm{obs}$] or [$R_\mathrm{obs} = \sqrt{x_\mathrm{obs}^2+y_\mathrm{obs}^2}$, $\phi_\mathrm{obs}$]) is obtained from the hit pattern of the S2 signal on the top PMT array. The origin in the $x$-$y$ plane is set to the center of the TPC. Several position reconstruction algorithms are employed in LXe detectors, such as artificial neural networks (NN)~\cite{xe100_analysis}, top pattern fit (TPF)~\cite{bartthesis}, support vector machine (SVM)~\cite{xe100_instru} and statistical light response functions (LRF)~\cite{lux_lrf}. In XENON1T, a NN is trained using the open-source Fast Artificial Neural Network Library (FANN)~\cite{fann}. In addition, a TPF algorithm serves as a cross-check for identifying events with poorly reconstructed positions (Sec.\,\ref{sec:Cuts}).

In order to calibrate the algorithms, data from an optical Monte Carlo (MC) simulation is used. Training data is generated by propagating photons over the full detector geometry~\cite{xe1t_mc} which is implemented using the GEANT4 toolkit~\cite{geant4}. Optical parameters such as the refractive index, PTFE (polytetrafluoroethylene, Teflon) reflectivity, xenon absorption length and Rayleigh scattering length are tuned by matching the simulated light collection to $^{83\mathrm{m}}$Kr calibration data.

\subsection{Field Distortion Correction}
\label{pos:cor}

%

\begin{figure}
\centering
\includegraphics[width=1\linewidth]{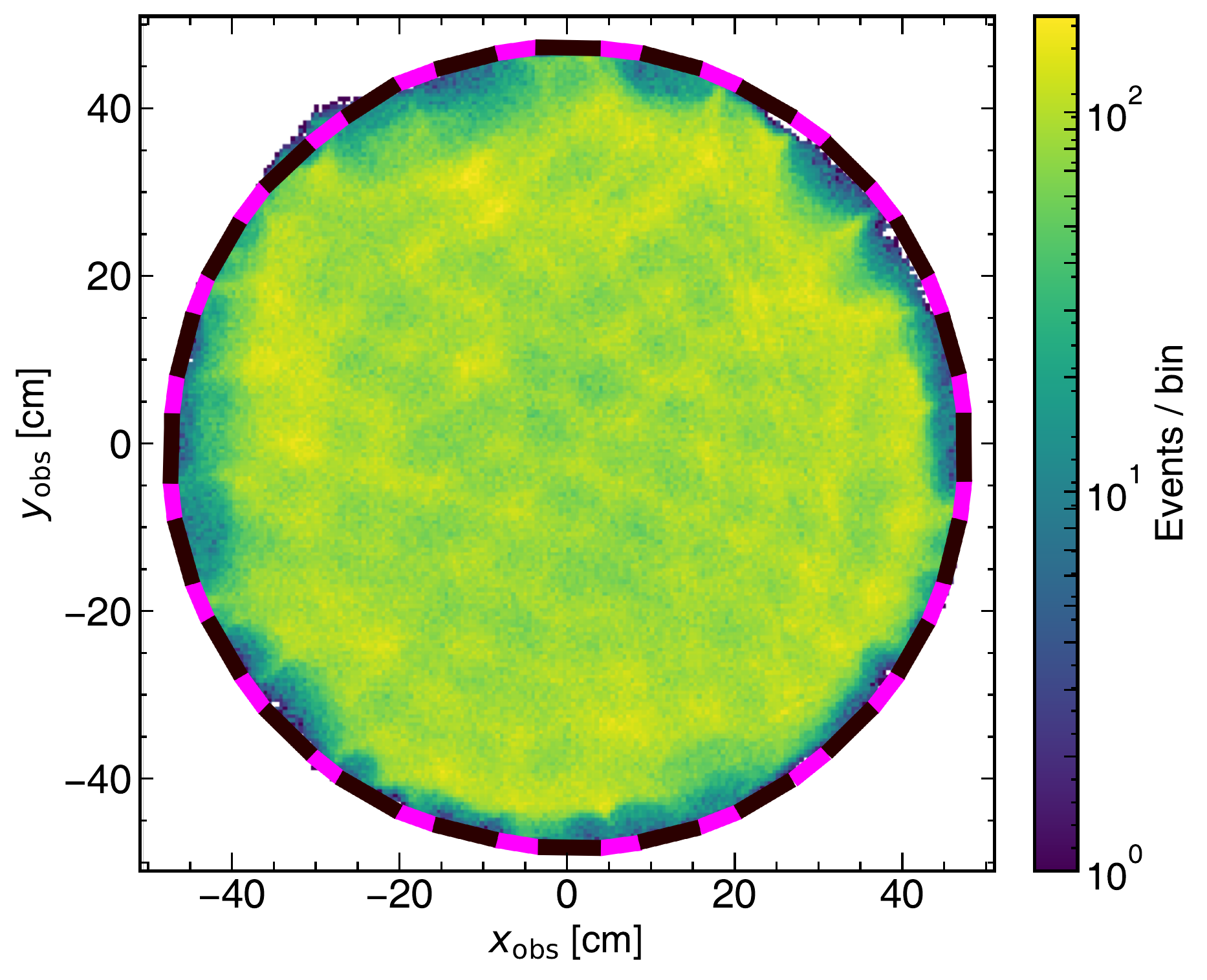}
\caption{$x_\mathrm{obs}$-$y_\mathrm{obs}$ distribution of $^{83\mathrm{m}}$Kr events as reconstructed by the FANN algorithm integrated over $z_\mathrm{obs}$. The distortions at high radii coincide with the 24 PTFE reflector panels (black segments) that are not in contact with the ring-shaped electrodes surrounding the TPC for drift field shaping. Magenta segments indicate the panels that are in contact with the electrodes.
}
\label{fig:bite_structure}
\end{figure}

\begin{figure}
\centering
\includegraphics[width=1\linewidth]{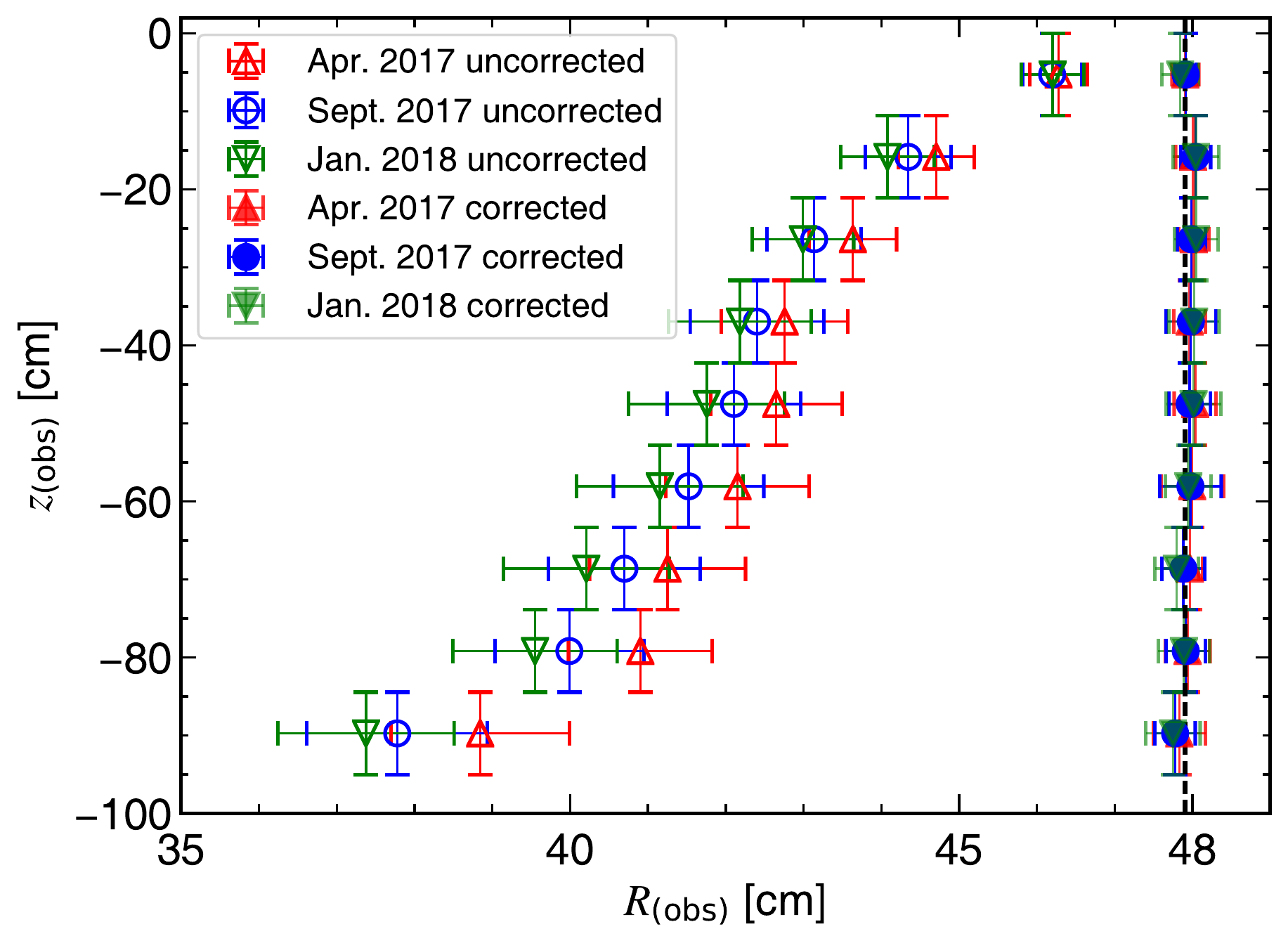}
\caption{Maximum TPC radius reconstructed from signals of surface events in three time intervals and in bins of $z$. Open (filled) markers show radii before (after) position correction. Horizontal error bars indicate the radial width of the event distribution. Vertical error bars mark the $z$ bin width. The black dashed vertical line indicates the geometrical TPC radius.}
\label{fig:bgcomp_timedp}
\end{figure}

Reconstructed spatial distributions exhibit a radial inward bias with increasing depth due to the distortion of the electric drift field and repulsive effects of accumulated negative charges on the lateral PTFE reflector panels that confine the TPC. Fig.\,\ref{fig:bite_structure} illustrates this effect integrated over the $z_\mathrm{obs}$ coordinate. The $x_\mathrm{obs}$-$y_\mathrm{obs}$ distribution for $^{83\mathrm{m}}$Kr data exhibits a regular geometric distortion from the physical TPC boundary that can be related to the PTFE panel configuration. The distortion is largest at the locations of the 24 panels (black segments) that are not in contact with the ring-shaped copper electrodes surrounding the TPC for drift field shaping~\cite{xe1t_instru}. The distortion is smallest at the smaller panels (magenta segments) which are in contact with the electrodes. Fig.\,\ref{fig:bgcomp_timedp} shows the position of the TPC edge in bins of $z_\mathrm{obs}$ by open markers for several periods throughout the science run, indicating an increasing accumulation of charges during detector operation. The data are derived from the radial distribution of signals from $^{222}$Rn progeny on the PTFE surface. Those signals are referred to as surface events~\cite{xe1t_long_analysis_2}. The error bars in $R_\mathrm{obs}$ direction indicate the event distribution's radial width while the error bars in $z_\mathrm{obs}$ mark the bin width.


The first WIMP dark matter results of XENON1T~\cite{xe1t_first} using data recorded in SR0, featured a two dimensional correction of reconstructed positions ($R_\mathrm{obs}$, $z_\mathrm{obs}$) based on a matching of the uniform spatial distribution of $^{83\mathrm{m}}$Kr events to the distribution predicted by electric field simulations performed with the COMSOL Multiphysics package~\cite{Comsol}. This correction is sufficient for a fiducial mass of up to $1\,$\ton since the contribution from surface events is negligible in the corresponding volume. During the analysis of SR1 data, an improved understanding of the field distortion and its time evolution was obtained which led to a $^{83\mathrm{m}}$Kr data-driven correction in three dimensions ($R_\mathrm{obs}$, $z_\mathrm{obs}$, $\phi_\mathrm{obs}$) for four time intervals throughout SR0 and SR1. To derive the correction, the detector is segmented into 180 bins in $\phi_\mathrm{obs}$ and 40 bins in $z_\mathrm{obs}$. The event positions in each bin are corrected such that they are evenly spaced in the square of the corrected radial position $R^2$. Corrected depths $z$ are subsequently obtained by the geometric relation $z = -\sqrt{z_\mathrm{obs}^2 - (R - R_\mathrm{obs})^2}$.

The position correction is validated with spatially homogeneous signals from decays of $^{131\mathrm{m}}$Xe activated by neutron calibrations as well as 
non-uniform sources, such as neutron calibration signals, which are compared to MC simulations. A Kolmogorov-Smirnov test \cite{Kolmogorov} yields a goodness of fit p-value of 0.55 when comparing $^{131\mathrm{m}}$Xe event positions to a uniform distribution.

The impact of the correction is illustrated in Fig.\,\ref{fig:bgcomp_timedp}. Filled markers indicate the corrected radial position of the TPC edge in bins of $z$.  After correction, the position of surface events coincides with the maximal radial position.
\subsection{Radial Position Resolution}
\label{pos_res}

The position resolution in the radial dimension is dependent on the S2 signal size as well as on the radial event position due to non-functional PMTs and light reflection at the TPC boundary.

\begin{figure}
\centering
\includegraphics[width=1.0\linewidth]{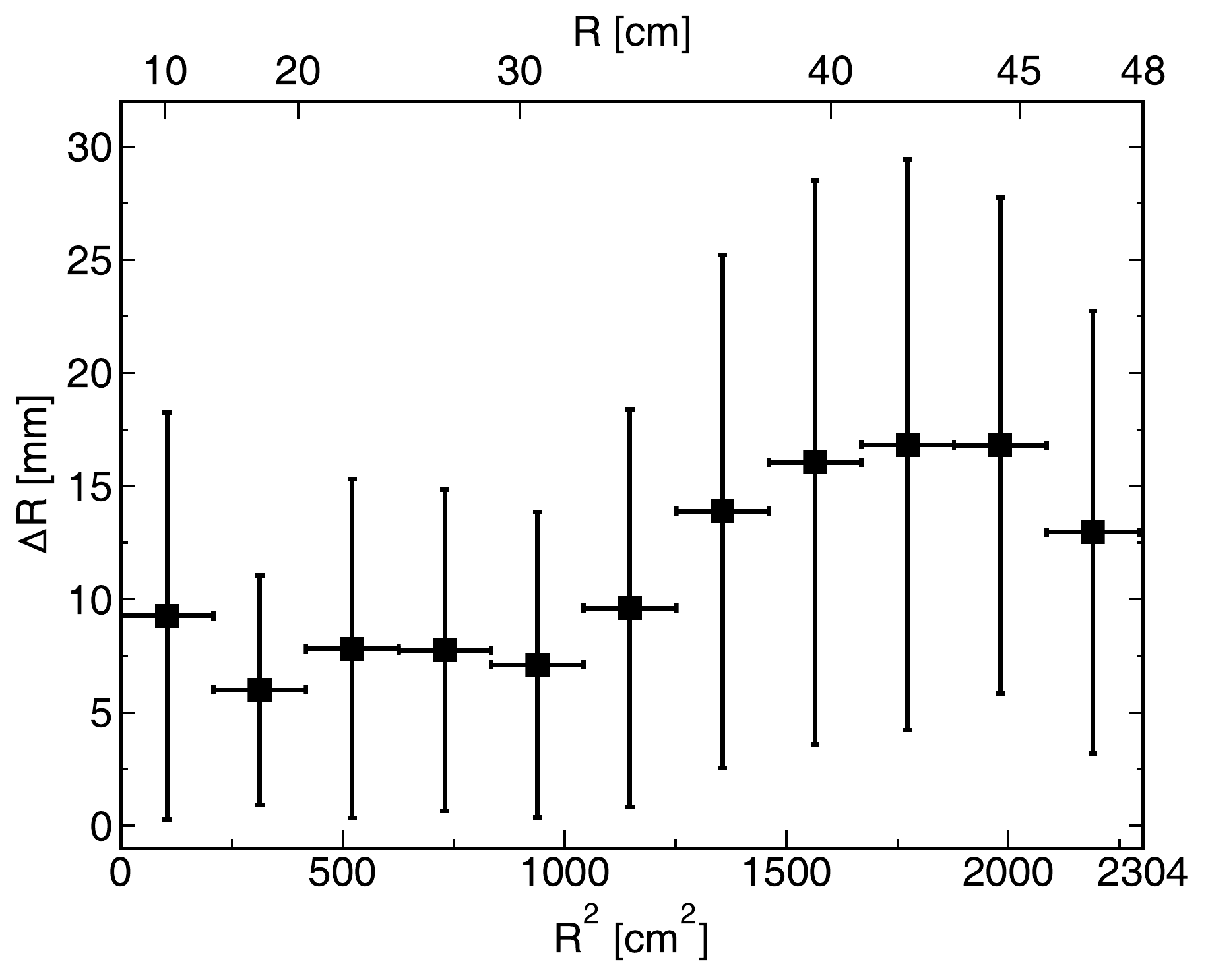}
\caption{Absolute radial difference ($\Delta R$) reconstructed between $32.1\,$keV and $9.4\,$keV signals from $^{83\mathrm{m}}$Kr in bins of $R^2$. The corresponding $R$ scale is shown on the upper horizontal axis. The vertical error bars represent the standard deviation of the $\Delta R$ distribution while horizontal error bars indicate the $R^2$ bin width.}
\label{pos_res_kr}
\end{figure}

The two consecutive $^{83\mathrm{m}}$Kr decays provide a sample to show the radial position resolution as a function of $R$ for fixed S2 signal sizes. 
Events in the upper $\sim5\,$cm of the TPC are selected as only in that region the two S2 signals can be resolved.
The spatial separation of the signals can be neglected with respect to the uncertainty from the reconstruction method. The average path length of the $9.4\,$keV conversion electrons in LXe is only $\sim$10\,$\mu$m and the small half-life of $157\,$ns does not allow for a reconstructable drift of the atoms by convection.
The mean of the distribution of the absolute radial difference ($\Delta R = |R_{32.1} - R_{9.4}|$) between the $32.1\,$keV and $9.4\,$keV signals is shown in bins of $R^2$ in 
Fig.\,\ref{pos_res_kr}.
The vertical error bars display the distribution's standard deviation. Note that the position reconstruction uncertainties from both decays are convoluted in $\Delta R$. Hence, a direct comparison to the resolution of single events is not possible.
While the precision in reconstructed radial positions is in the order of $\sim$1\,cm for $R<35\,$cm, the performance declines by a factor of 1.5 towards larger radii. This is caused by non-functional PMTs and light reflection at the TPC boundary as mentioned above. 

\begin{figure}
\centering
\includegraphics[width=1.0\linewidth]{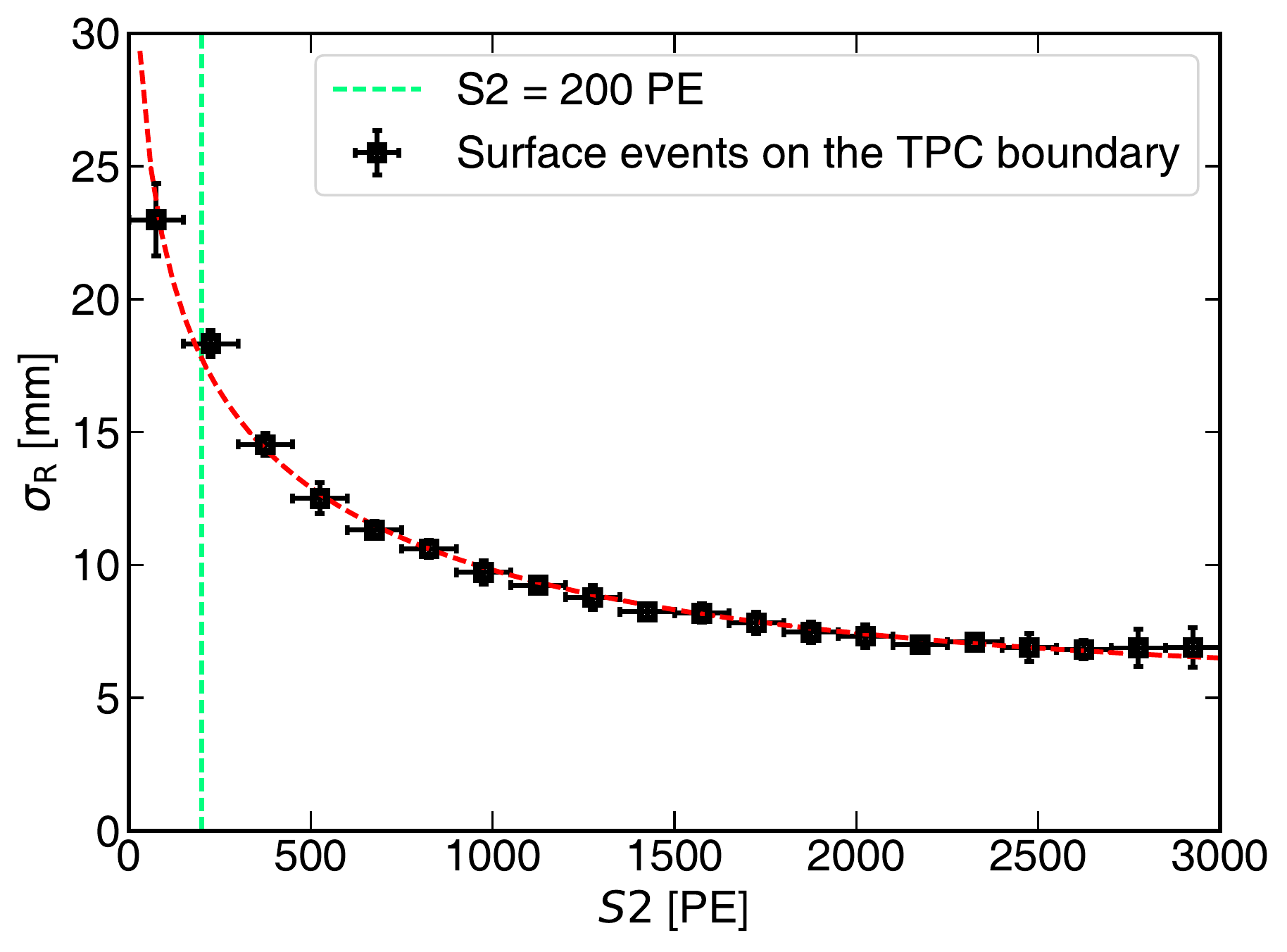}
\caption{Radial position resolution $\sigma_R$ in bins of S2 signal size for surface events. Vertical uncertainties are derived from the Gaussian fit while horizontal error bars mark the
S2 bin width. The red dashed line indicates the best-fit of an empirical function.}
\label{pos_res_surface}
\end{figure}

In addition to the radial dependence of the position resolution for fixed S2 signal sizes we also investigated its dependence on S2 signal sizes for a fixed position.
Surface events provide a sample at the maximum TPC radius and cover a large range of S2 signals down to less than 200\,PE due to charge loss.
Fig.\,\ref{pos_res_surface} shows the standard deviation $\sigma_R$ of the radial distribution in bins of S2 signal size. The uncertainties are derived from the gaussian fit and horizontal error bars mark the S2 bin width. $\sigma_R$ yields $1.9\,$cm at the S2 analysis threshold of $200\,$PE and decreases to values $<0.8\,$cm for large S2 signal sizes. The resolution is limited by the accuracy of the optical MC simulation used to train the FANN. Note that the resolution is poorest for surface events due to their location at high radii (Fig.\,\ref{pos_res_kr}) and is expected to improve for interactions in the center of the TPC.  

The presented position resolution studies are considered in the background and signal models for final inference of the dark matter search data~\cite{xe1t_long_analysis_2}.

%% file: SignalCorrections.tex
\section{Signal Corrections}
\label{sec:Sign_Corr}


The size of the recorded S1 and S2 signals depends on the event location in the detector due to various position-dependent effects, such as electron attachment to impurities in the LXe target, light collection efficiency, field inhomogeneities, variations of the thickness of the proportional scintillation region and non-functioning PMTs.
In the following, the corrections applied to S1 and S2 signals in order to account for these effects are explained. The corrected signals are denoted as cS1 and cS2.

\subsection{Electron Lifetime Correction}
\label{sec:ELifetimeCorr}

\begin{figure*}
\centering
\includegraphics[width=\textwidth]{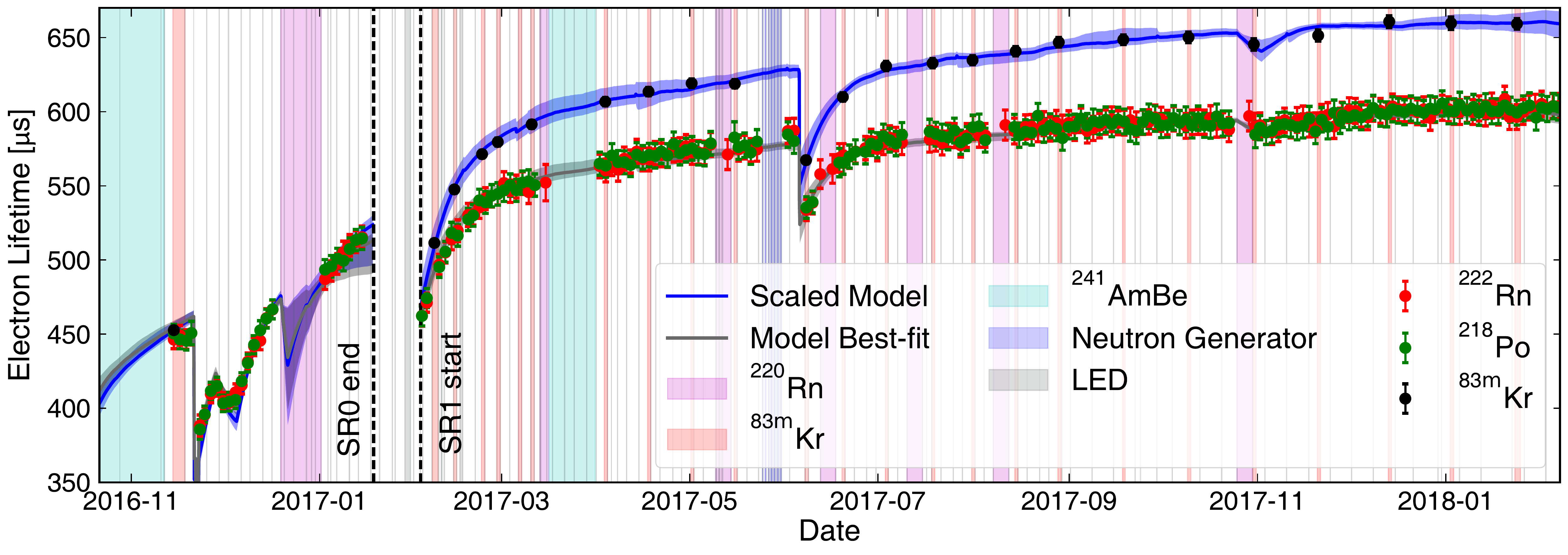}
\caption{Electron lifetime evolution during the two science runs measured from $^{83\mathrm{m}}$Kr (black), $^{222}$Rn (red) and $^{218}$Po (green) decays. Decreases are caused by releases of impurities due to changes in detector operation parameters like the detector's cooling power and the gas flow in the purification system. The temporal fine structure is modeled based on the $\alpha$ measurements (grey line) while the absolute scale of the electron lifetime is determined from the $^{83\mathrm{m}}$Kr measurement.}
\label{fig:elifetime_evolution}
\end{figure*}

The loss of ionization electrons due to attachment to electronegative impurities (e.g. O$_2$)~\cite{Bakale_1976} in LXe is a function of the drift time and follows an exponential law with the electron lifetime $\tau_e$ as a decay parameter.
This effect is the most important correction for S2 signals and is highly dependent on the impurity concentration in the target.
Since the xenon is continuously purified, $\tau_e$ is a parameter changing over time and has to be continuously monitored~\cite{xe1t_instru}.

The electron lifetime is evaluated in intervals of two to three weeks by measuring the $41.5\,$keV signal from the two consecutive $^{83\mathrm{m}}$Kr decays as a function of electron drift time. Additionally, $\tau_e$ is estimated from mono-energetic $\alpha$-decays of $^{222}$Rn and $^{218}$Po observed in background data that provide sufficient statistics on a daily basis. The $\tau_e$ values from these two methods are shown in Fig.\,\ref{fig:elifetime_evolution} and feature an offset of up to 10\% of unknown origin that scales with the xenon purity. 
The best hypothesis of the discrepancy is related to the small inhomogeneity of the drift field. The ionization yield of $\alpha$-decays has stronger field dependence than ERs and NRs, hence the measured $\tau_e$ are different. Because the energy from $^{83\mathrm{m}}$Kr decays is closer to the region of interest for dark matter searches compared to $\alpha$-decays, furthermore, better energy resolution up to the MeV scale and better discrimination between ER and NR signals~(Sec.\,\ref{sec:Discrimination}) can be achieved when applying $\tau_e$ derived from $^{83\mathrm{m}}$Kr decays in the S2 signal correction, we decided to use the $\tau_e$ from $^{83\mathrm{m}}$Kr decays for the final corrections.


The temporal fine structure of the electron lifetime evolution is modeled based on $\alpha$-decays by 
fitting a model that estimates the evolution of impurity concentrations in the gaseous and LXe phase and takes into account various detector operation parameters like the detector's cooling power and the xenon gas flow in the purification system \cite{xe1t_purification}. The model's best-fit (uncertainty) is shown by the grey line (band) in Fig.\,\ref{fig:elifetime_evolution}. During the two science runs of XENON1T, several decreases of electron lifetime were observed that coincide with releases of impurities due to changes in the above mentioned parameters.

The absolute scale of the electron lifetime model is derived by relating the 1/$\tau_e$ data points from the two methods by a linear function which is used to scale from the $\alpha$ measurement to the $^{83\mathrm{m}}$Kr measurement. The final electron lifetime model used for S2 signal correction is shown by the blue solid line together with its uncertainty band.

 During SR1, $\tau_e$ leveled off at about $650\,\mu$s corresponding to an oxygen equivalent impurity concentration of about $0.5\,$ppb limited by outgassing materials and the flow in the gas purification circuit.


\subsection{S2 Spatial Correction}
\label{sec:Sign_Corr:S2_amp}


 The proportional scintillation signal S2 is produced
 between the liquid-gas interface and the anode electrode. 
 The fraction of the total S2 signal measured by the top PMT array is $(63\pm2)\,$\%. The signal detected in the top array is highly localized, while the bottom PMT array provides a more uniform distribution that is more resilient to effects from non-functional PMTs or variable light collection efficiency.
 For this reason, only the corrected bottom array signal, cS2$_\mathrm{b}$, is used in the final inference of dark matter search data.
 
 Positional variations in the S2 size arise due to distortion of the electric field at the liquid-gas interface induced by subsidence of the anode caused by its weight, impacting the electron extraction efficiency. These variations are corrected using the $41.5\,$keV charge signal of $^{83\mathrm{m}}$Kr.
The $x_\mathrm{obs}$-$y_\mathrm{obs}$ distribution of S2 signals is fit by a 2D second-order polynomial.
 The best-fit value of the function's center for S2 signals observed by the bottom array
 is displaced from the origin by about $1.5\,$cm to negative $x_\mathrm{obs}$ and $y_\mathrm{obs}$ values.
 This displacement indicates a slight tilt of the TPC.
 The extraction efficiency is approximately 20\%-30\% lower at the edge of the detector compared to the center while the average value yields 96\% \cite{xe1t_instru}.

\subsection{Light Collection Efficiency}


 \begin{figure*}
    \centering
    \includegraphics[width = 0.45\textwidth]{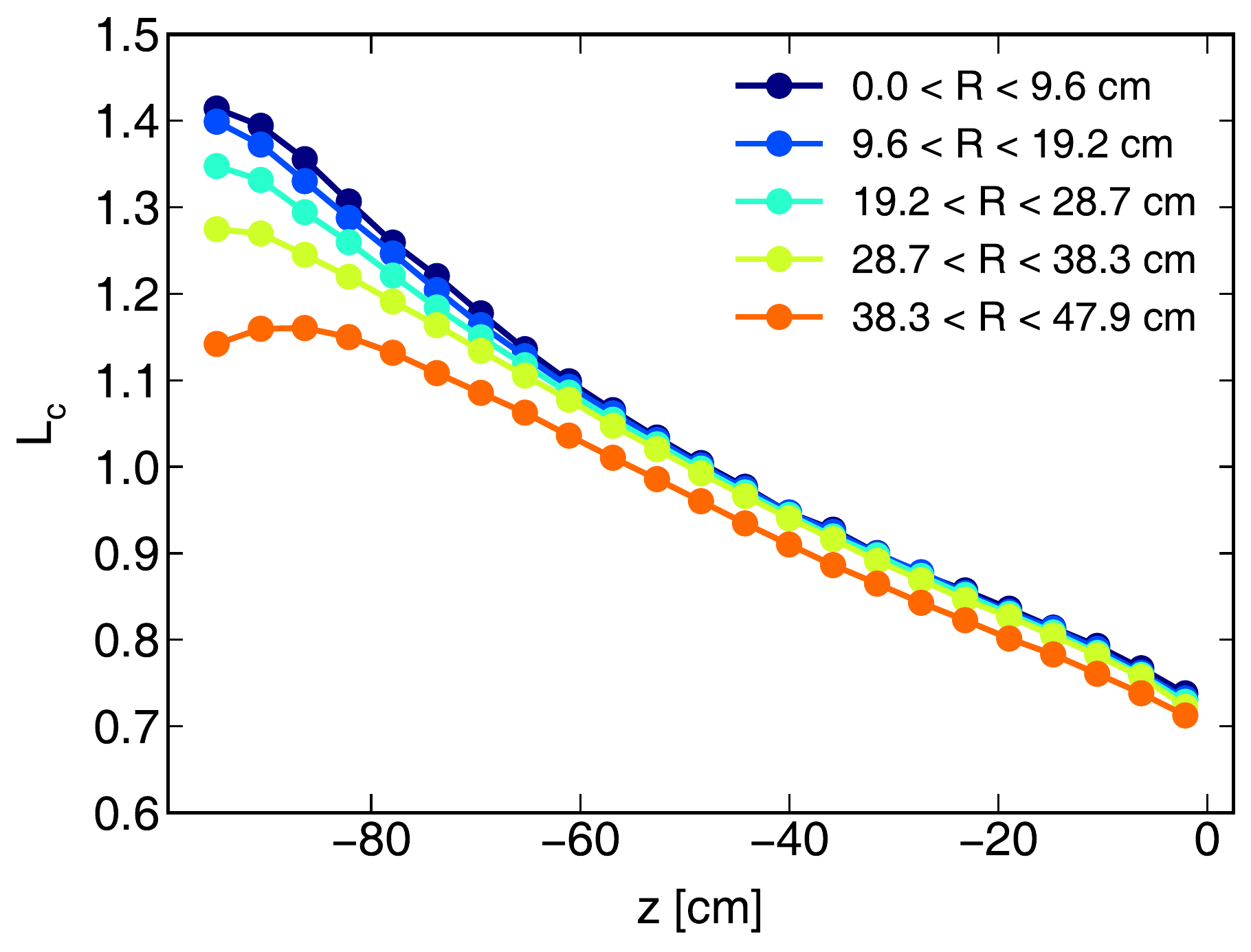}
    \includegraphics[width = 0.465\textwidth]{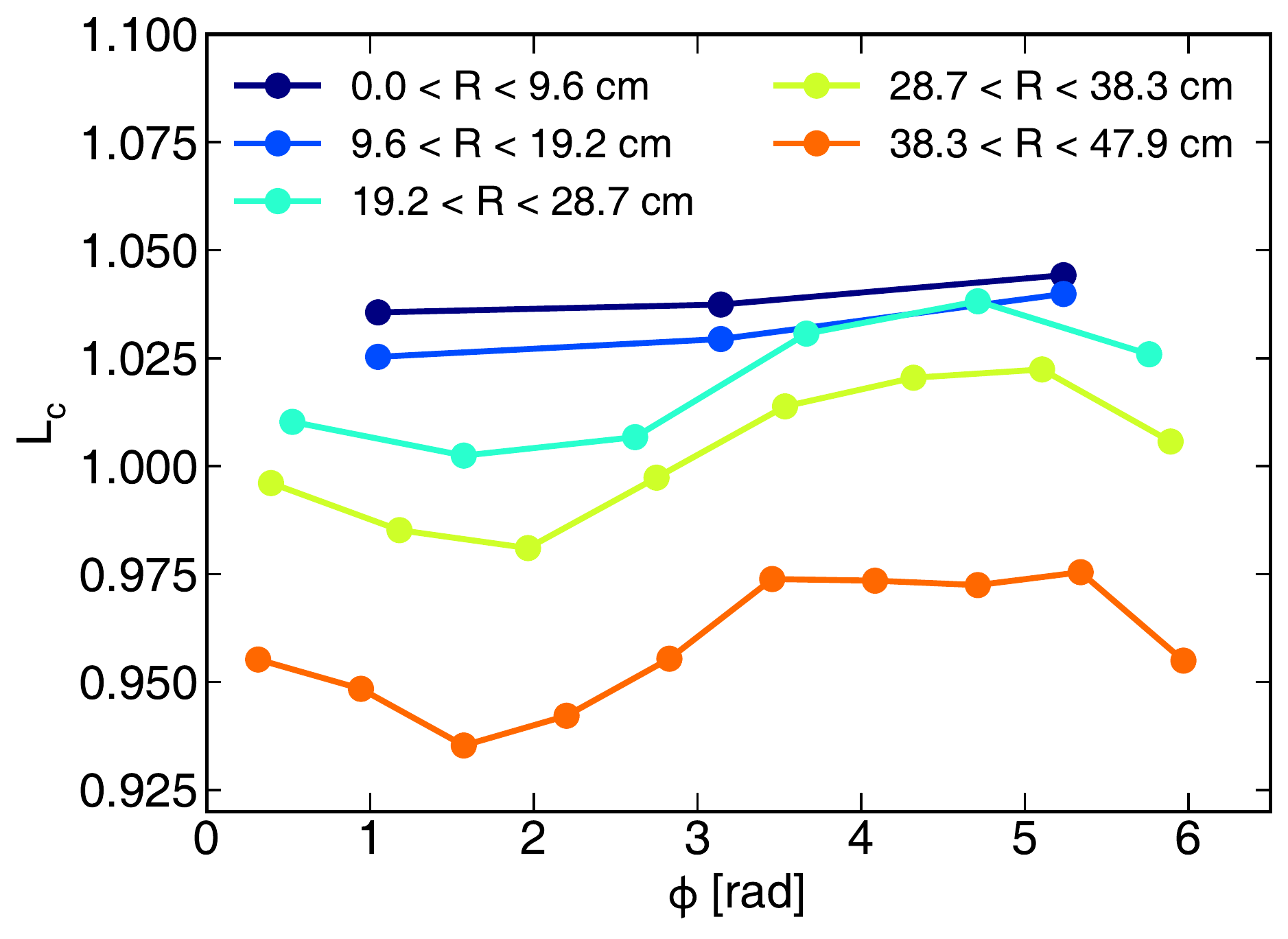}
    \caption{Spatial dependence of the relative light collection efficiency $L_c$ 
    on $z$ (left) and $\phi$ (right) for different radial bins (color code). Data points are connected by straight lines to guide the eye.}
    \label{fig:lce_variations}
\end{figure*}

The LY is impacted by the light collection efficiency $\epsilon_L$, i.e., the number of photons that hit a PMT photocathode per photon emitted at the interaction site, the photon yield (PY), i.e., the number of generated photons per incident energy $\varepsilon$, the PMT quantum efficiency $\epsilon_\mathrm{QE}$, i.e. the probability that one photon hitting the photocathode induces one PE, and the collection efficiency $\epsilon_\mathrm{CE}$ of photo-electrons within the PMT:
\begin{equation}
\label{equ:ly}
 \begin{aligned}
    &\frac{\mathrm{S1}(R,\phi,z,\varepsilon,F)}{\varepsilon} = \:
     \mathrm{LY}(R,\phi,z,\varepsilon,F) \\
     &= \epsilon_L(R,\phi,z) \cdot \mathrm{PY}(\varepsilon,F(R,\phi,z,t))\cdot \epsilon_\mathrm{QE}\cdot\epsilon_\mathrm{CE}.
  \end{aligned}
\end{equation}
$\epsilon_L$ is affected by the number of photon reflections before reaching a PMT photocathode and is, therefore, spatially dependent. The PY depends on the energy and the drift field $F$ which both impact on the electron-ion recombination~\cite{NEST2.0}. The field features variations at the TPC edges which vary in time $t$ due to charge accumulation on PTFE surfaces (see \secref \ref{sec:ProRec}).

In order to correct for the spatial dependence of S1 signals, a three-dimensional correction map is derived from the $32.1\,$keV signals in $^{83\mathrm{m}}$Kr calibration data. 
The mean of the S1 distribution is evaluated in discrete ($R$, $\phi$)-regions and in slices of $z$, and is normalized to its average $\langle\mathrm{S1}\rangle$ across the TPC in order to obtain the relative light collection efficiency $L_c$ that removes the spatial dependencies:
\begin{equation}
\begin{aligned}
    &\frac{LY(R,\phi,z,\varepsilon,F)}{L_c(R,\phi,z)} \:= \langle LY(\varepsilon, F) \rangle \\&= \langle \epsilon_L \rangle \cdot \langle PY(\varepsilon,F(R,\phi,z,t)) \rangle \cdot \epsilon_\mathrm{QE} \cdot \epsilon_\mathrm{CE}.
    \end{aligned}
\end{equation}

The number of bins of the correction map was optimized in each dimension by limiting the maximum variation to be about 2.5\% between two adjacent bins.
The correction not only averages out the spatial dependence of $\epsilon_L$ but also accounts for the spatially dependent $PY$ introduced by field inhomogenities.

$^{83\mathrm{m}}$Kr decay energies lie beyond the region of interest for WIMP searches and lower energetic events are less sensitive to changes in the field.
Hence, a small bias of $\sim2$\% \cite{NEST2.0} is introduced when applying $L_c$ to the WIMP search region and $L_c$ varies in time by up to 6\% due to evolving field inhomogenities (Sec.\,\ref{pos:cor}).
To remove this bias, the spatial distribution of the CY from the $41.5\,$keV $^{83\mathrm{m}}$Kr signal is used to map out
local and timely field variations and decouple those from both, the $L_c$ and electron lifetime corrections. Since the CY is correlated with the electron lifetime, this procedure is repeated iteratively until convergence is observed resulting in a time stability of $L_c$ within $1.2$\%.

Fig.\,\ref{fig:lce_variations} shows $L_c$ measured as a function of $z$ (left) and $\phi$ (right) for bins in $R$. The largest variation is observed along $z$ with a maximum $L_c$ at the bottom center of the detector where the solid angle to the bottom PMT array is largest. 



%% file: Cuts.tex
\section{Selection criteria and their acceptances}
\label{sec:Cuts}


This section describes the criteria applied to the dark matter search data for selecting single scatter events in the region confined by cS1$\in[3,70]\,$PE and cS2$_\mathrm{b}\in[50, 7940]\,$PE correspond to the energy region of interest, [1.4, 10.6]\,keV$_\mathrm{ee}$ ([4.9, 40.9]\,keV$_\mathrm{nr}$)~\cite{xe1t_comb}. Note that the cS2$_\mathrm{b}$ was used in the analysis due to its uniform distribution as explained in Sec.\,\ref{sec:Sign_Corr:S2_amp}. 50\,PE of cS2$_\mathrm{b}$ corresponding to 100\% acceptance for the events with S2 over its 200\,PE trigger threshold.

\subsection{Data Quality Selection}
\label{sec:cuts:LivetimeReduction}

Operational conditions during data acquisition are necessary for the rejection of certain time periods regardless of the properties of the events contained within. The corrected live time and respective acceptance after the incremental application of four criteria are summarized in Tab.\,\ref{tab:LivetimeReduction}.

The \textit{DAQ veto} ensures that all channels in the DAQ system are able to record data. If this is not the case, a system-wide busy condition is issued. The start and stop times of the busy signal are saved in the data stream ensuring that those time periods can be removed at analysis level. The DAQ veto rejects about 6\% (1\%) of data in SR0 (SR1), with the increased deadtime in SR0 caused  by non-optimized DAQ settings during the first XENON1T runs.

The active Cherenkov \textit{Muon veto} triggers if at least eight PMTs in the water tank record signals larger than $1\,$PE within a $300\,$ns time window. Under these conditions, a muon tagging efficiency of 99.5\% is achieved while muon-induced shower events are identified with a probability of 43\%. These efficiencies are determined from simulations and the events are required to exhibit at least one produced neutron that has high enough energy ($>10\,$MeV) to reach the TPC~\cite{xe1t_muonveto}. To remove signals from potential secondary interactions in the TPC, a muon veto trigger is searched in a window of [$-2\,$ms, $+3\,$ms] around each TPC event which is rejected in case of success. The time range is conservatively determined from simulations~\cite{xe1t_muonveto} and data.
In addition, all data are removed where the muon veto is inactive. In total about 2\% (1\%) of live time after the DAQ veto is removed in SR0 (SR1) due to the muon veto criterion. The muon veto reduces the expected muon-induced neutron background rate by a factor of~2.5.

The XENON1T PMTs can emit bursts of light as previously observed in independent measurements \cite{PMTs_xe1t}.
In SR0 (SR1) a total of 8 (179) light flashes were observed, causing short periods of high pulse rates throughout the TPC from both the primary light and secondary interactions. These incidents are removed from the data by a \textit{Flash veto} that scans the pulse rates for each PMT channel and identifies sudden, drastic increases. A conservative time window of $10\,$s before and $120\,$s after each flash is rejected. After the application of the DAQ and muon veto, the flash veto removes 0.04\% and 0.12\% of livetime in SR0 and SR1, respectively. Flashes trigger in most cases the DAQ busy signal. Hence, the criterion is highly correlated with the reduction of live time due to the DAQ veto.

\begin{table}
\centering
\begin{tabular}{|l|cc|cc|}
\hline
 \multicolumn{1}{|c|}{Data quality} &
\multicolumn{2}{|c|}{Live time [days]} & \multicolumn{2}{|c|}{Incremental Acceptance [\%]} \\
\multicolumn{1}{|c|}{criterion}& SR0 & SR1 & SR0 & SR1\\\hline
Without Cut & 37.2 & 264.8 & 100 & 100\\
DAQ veto & 34.2 & 261.6 & 92.1 & 98.8\\
Muon veto & 33.5 & 259.1 & 90.0 & 97.8\\
Flash veto & 33.5 & 258.8 & 89.9 & 97.7\\
S2 tails & 32.1 & 246.7 & 85.8 & 93.0\\\hline
\end{tabular}
    \caption{Summary of data live time and respective acceptance after incrementally applying data quality requirements in the shown order.}
    \label{tab:LivetimeReduction}
\end{table}

\begin{figure}
    \centering
    \includegraphics[width = 0.49\textwidth]{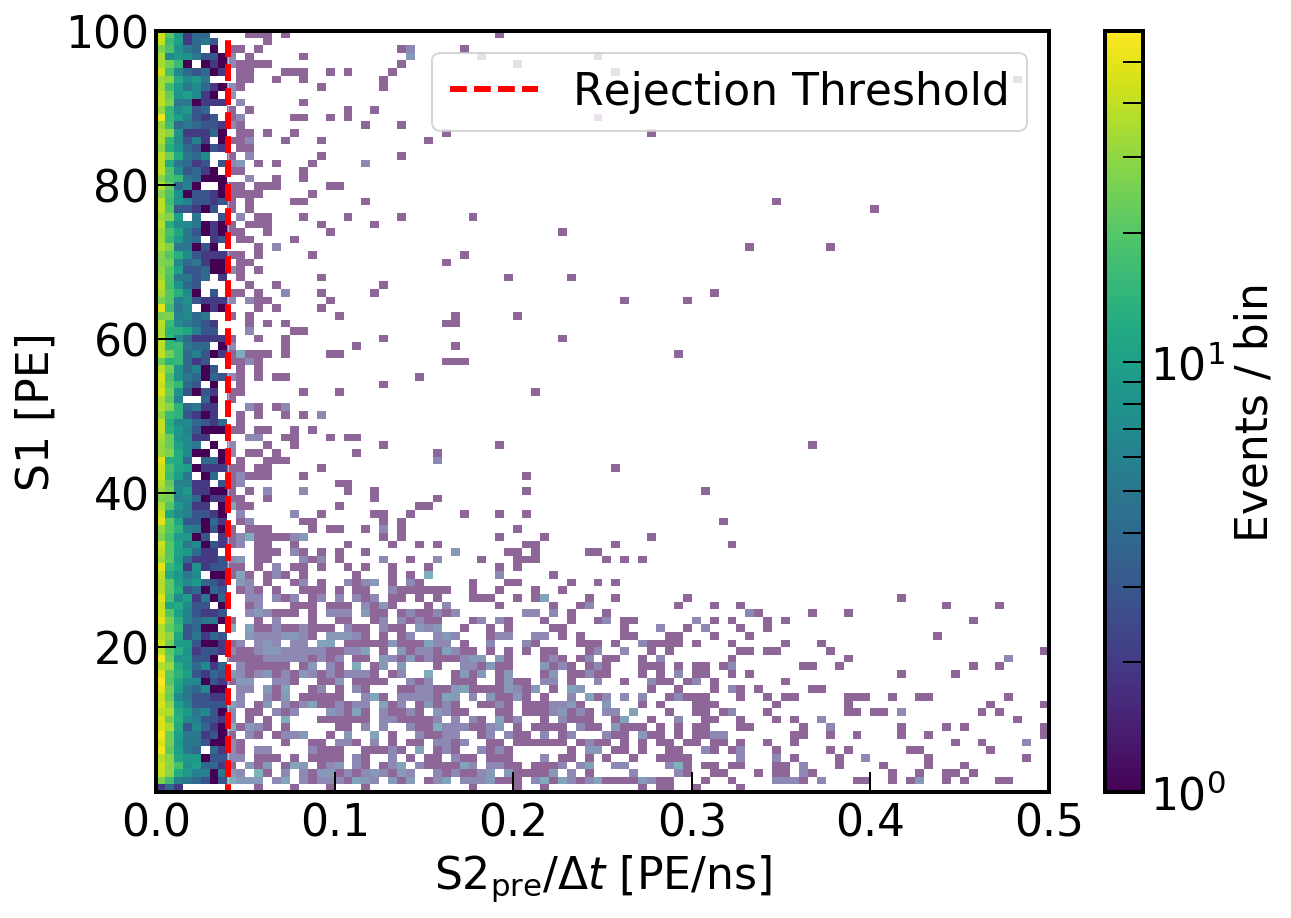}
    \caption{Distribution of S1 signal sizes vs. $\mathrm{S2}_\mathrm{pre}/\Delta t$ as measured in $^{220}$Rn calibration data. The threshold in $\mathrm{S2}_\mathrm{pre}/\Delta t$ above which events are rejected regardless of their own properties is indicated by the red line.}
    \label{fig:cuts:S2TailsPlot}
\end{figure}

The \textit{S2 tails} criterion addresses delayed S2 signals, e.g. from delayed electron extraction, or photo-ionization on materials and impurities, that are generated especially after large S2s and do not correspond to physical interactions. Those can reduce the detector sensitivity to low-energy interactions for several ms. For each event, the discrimination variable $\mathrm{S2}_\mathrm{pre}/\Delta t$ is defined as the ratio of the S2 size of a preceding event divided by the time difference to that event, where the preceding 100 events are scanned and the maximum of the parameter is stored. Fig.\,\ref{fig:cuts:S2TailsPlot} shows the distribution of primary S1 signal sizes vs. $\mathrm{S2}_\mathrm{pre}/\Delta t$ in single scatter events (Sec.\,\ref{sec:cuts:singlescatter}) of background data while a sub-set of noise rejection and reconstruction requirements (Sec.\,\ref{sec:cuts:reconstruction}) has been applied as a pre-selection. Ionization signals from preceding events can be mis-identified as an S1 or S2 of an interaction in subsequent events and therefore appear as a horizontal population in the figure. The $\mathrm{S2}_\mathrm{pre}/\Delta t$ threshold above which an event is rejected is set to $0.04\,$PE/ns, which is chosen to remove the most intense region of increased activity while maintaining as much live time as possible.
This removes 4\% (5\%) of live time in SR0 (SR1) in addition to the previously-mentioned live time reductions.

\subsection{Noise Rejection and Reconstruction Requirements}
\label{sec:cuts:reconstruction}


A set of conditions is imposed to remove events that are either falsely reconstructed, members of known background populations, or generally of low quality:
\begin{itemize}
    \item If an event contains a large integral ($>$\,$300\,$PE) of signals prior to the primary S2 excepting the primary S1, the event is deemed ``noisy'' and is removed.
    \item The contribution of one channel to an event's S1 is not allowed to exceed 5\% of the S1 plus an off-set of four PE. This criterion prohibits that a single channel exhibiting a PMT malfunction dominates the signal. Typical causes for failing the condition are PMT afterpulses or light emission.
    \item S2s originating from single electron signals can be mis-classified as S1s at the data processor level. This can result in events where two S2 signals, one mis-identified and one lone signal, are randomly paired as an interaction. Lone S2 signals originate from delayed electron extraction and pile-up and therefore do not feature a corresponding S1. Two machine learning classification algorithms, a boosted decision tree and a random forest from the scikit-learn python package \cite{scikit-learn}, are employed to reclassify S1s based on the most important peak properties width, area, rise time and signal fraction detected in the top PMT array. Training samples for good S1s are selected from high-quality background events in the ER band and from sampling hits from larger S1 signals, effectively creating smaller signals in the region of interest. Single electron S2 training samples are created by selecting S2 peaks that are isolated from other signals by at least $10\,\mu$s. The threshold for removing an event, placed on the classifier's normalized voting, was optimized to achieve a reduction of S2 signals mis-classified as S1 signals by a factor of 5.
    \item The almost constant fraction of light from S2 signals observed by the top PMT array is used to reject background caused by interactions in the gas phase above the anode electrode or from mis-reconstructed events.
High-quality calibration events are used to model the distribution of true interactions in the liquid xenon depending on the S2 size. Events that exhibit an S2 light fraction in the top array that is smaller or larger than the 99\% quantiles of the distribution are rejected.
\item Reliable position reconstruction is ensured by demanding the reconstructed $x_\mathrm{obs}$-$y_\mathrm{obs}$-coordinates to be consistent with the S2 hit pattern on the top PMT array. This criterion predominantly suppresses pile-up of delayed electron signals, double scatters or events that are mis-reconstructed at the wrong $x_\mathrm{obs}$-$y_\mathrm{obs}$-position, often due to non-functional PMT channels. The likelihood of the observed pattern given the position is computed using the same optical MC simulation as employed for the training of the position reconstruction algorithms. 
\item To further suppress anomalous $x_\mathrm{obs}$-$y_\mathrm{obs}$ reconstruction, events are removed if the difference in the reconstructed positions for the two reconstruction algorithms exceeds the upper 99\% quantile of the position difference distribution defined in dependence of S2 size. This distribution is extracted from high-quality calibration data. 
\end{itemize}


The acceptance of each cut described above is evaluated individually based on control samples either from calibration or background data. All data quality criteria together accept $(95\pm 4)$\% of true signals that fall into the region of interest.

\subsection{S1-S2 Signal Correlation Requirements}

\begin{figure}
    \centering
    \includegraphics[width=1\linewidth]{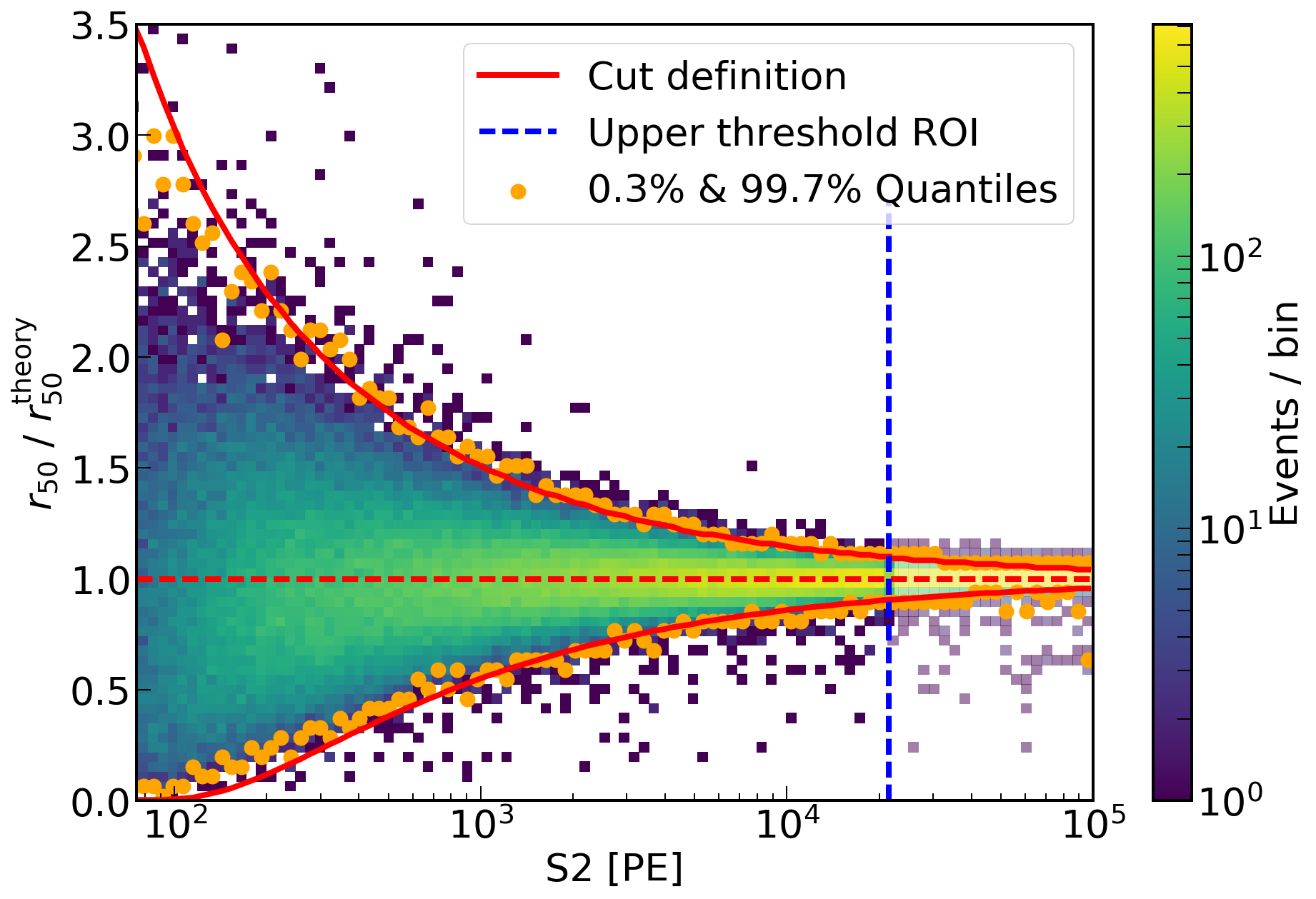}
    \includegraphics[width=1\linewidth]{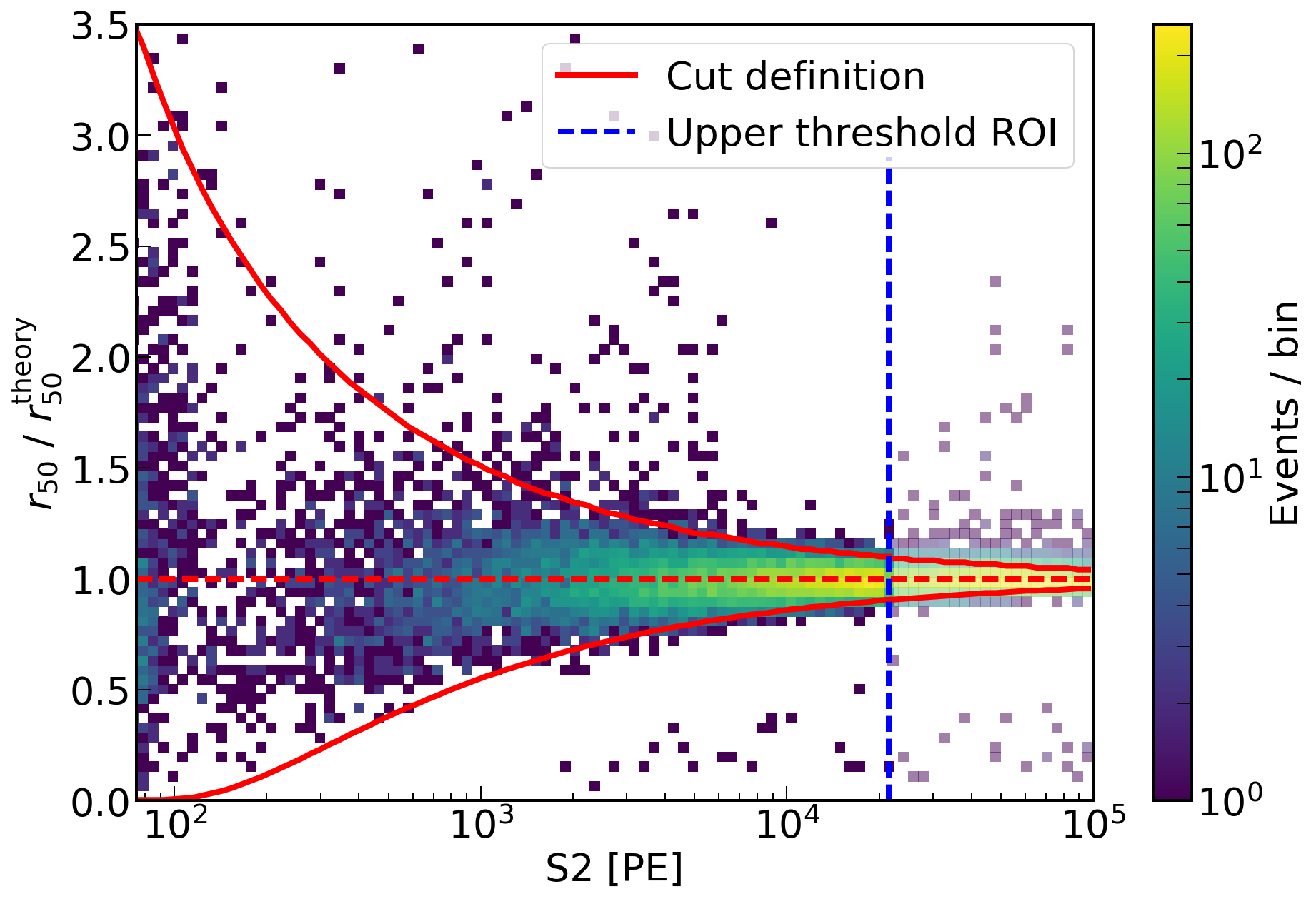}\\
    \caption{Distribution of the measured S2 width parameter $r_{50}$ normalized to the width $r^\mathrm{theory}_{50}$ expected from diffusion as function of S2 signal size for simulated (top) and combined $^{220}$Rn and $^{241}$AmBe calibration data (bottom). Yellow points mark the 0.3\% and 99.7\% quantiles for simulated data and red solid lines show the chosen cut definition. The red dashed line marks $r_{50}$ = $r^\mathrm{theory}_{50}$.}
    \label{fig:cuts:S2width}
\end{figure}

In each time window that is allocated to an event, the processor considers the largest S2 candidate and the S1 with the largest PMT coincidence level before the S2 as the primary interaction.
This and the following criteria suppress pile-up effects, double scatters, or accidentally pairing lone S1 and S2 signals that arise in charge- and light-insensitive detector regions:

\begin{itemize}
\item The width of the S2 peak is required to be correlated with the event's drift-time $t_\mathrm{d}$ and S2 signal size as expected from physical interactions due to diffusion of the electron cloud. The time interval $r_{50}$ in which 50\% of the S2 peak is contained can be modeled by \cite{diffusion_model}:
\begin{equation}\label{eq:3}
    r_{50}^\mathrm{theory} = \sqrt{\frac{3.64\cdot D\cdot t_\mathrm{d}}{v_\mathrm{d}^2}+w_0^2},
\end{equation}
with the free parameters drift velocity $v_\mathrm{d}$, diffusion constant $D$ and an adjustable constant $w_0$. In case of a Gaussian SE response, $w_0$ would be equal to the width of very shallow S2s. The factor 3.64 adjusts the diffusion model, originally defined for S2 signal widths of one standard deviation, to the 50\% width $r_{50}$. The three parameters are determined from calibration data and the model is implemented in the waveform generator. 
Using simulated waveforms, a large data set that extends across the full S2 signal size spectrum, which is not entirely covered by the calibration data, is generated. Figure.\,\ref{fig:cuts:S2width} shows the measured $r_{50}$ distribution normalized to $r_{50}^\mathrm{theory}$ as a function of S2 signal size for simulated (top) and calibration (bottom) data. The calibration data includes both $^{220}$Rn and $^{241}$AmBe data in order to provide statistics at low as well as at high S2 signal sizes. The broadening of the distribution for small S2 signal sizes is caused by binomial fluctuations of the number of electrons contributing to the signal. The S2 signal width requirement is constructed based on the quantiles (yellow points in Fig.\,\ref{fig:cuts:S2width}) in simulated data such that physical interactions are accepted in more than 99\% of the cases. The selection criterion derived from the quantiles is marked by the red lines and events not contained in between are rejected.
\item The PMT hit pattern of the S1 signal depends on the event position due to geometrical effects. Because the $x$-$y$ position is extracted from the S2 signal and the $z$ position is computed from the drift time, compatibility of the S1 signal's PMT pattern (S1 pattern) with this position represents an independent confirmation of the interaction pairing. The likelihood for an S1 pattern to originate from the reconstructed event position is computed by means of optical MC simulations. The selection criterion is tuned such that S1-S2 signal pairs from physical interactions defined by calibration control samples are accepted with a probability larger than 99\%.
\item Similarly, the fraction of the S1 signal detected by the top PMT array is required to be correlated to the interaction depth due to geometrical effects. The probability of a photon being detected in the top array is evaluated for each position in the target volume by means of optical MC simulation. This number is used to construct a binomial distribution from the total number of photons detected in both PMT arrays for a given event. If the actual fraction of photons detected in the top array falls into the extreme tails of the binomial distribution ($p < 0.001$), the event is removed. 
\end{itemize}

The described S1-S2 signal correlation criteria has a roughly flat acceptance in the region of interest (Fig.\,\ref{fig:cuts:acceptances}) with an average value of $(96\pm 3)$\% that estimated from control samples in background and calibration data.

\subsection{Single Scatter Requirements}
\label{sec:cuts:singlescatter}

Given the small expected scattering cross-section of dark matter particles, the probability that a WIMP scatters more than once in the TPC target volume is negligible. Hence, the identification of multiple scatter events is a powerful discriminator between signal candidates and background from radiogenic neutrons that induce identifiable multi-site events with a probability larger than 80\%.
\begin{itemize}
    \item In the S1 channel, events where a second S1 signal is accidentally present in the waveform are identified by searching for an additional S1 signal within one maximal drift time interval, i.e. the time interval that corresponds to the height of the TPC, before the primary S2. The drift time of this constructed alternative interaction is tested for a correlation with the primary S2 by using the S2 width selection criterion described in the previous section. If the S2 signal is compatible with an interaction at the depth indicated by this alternate pairing, the event is removed.
    \item In the S2 channel, events are removed if a second S2 of sufficient size is identified, with the size threshold defined as a function of the primary S2. The requirement results in a multiple scatter rejection efficiency of about 96\% and a single scatter acceptance of 99\% as estimated from simulations and NR calibrations.
\end{itemize}

The described criteria accept more than $99$\% of true single scatter signals that fall into the region of interest.

\subsection{$^{220}$Rn Calibration Data Specific Criteria}

\begin{figure}
    \centering
    \includegraphics[width = 0.49\textwidth]{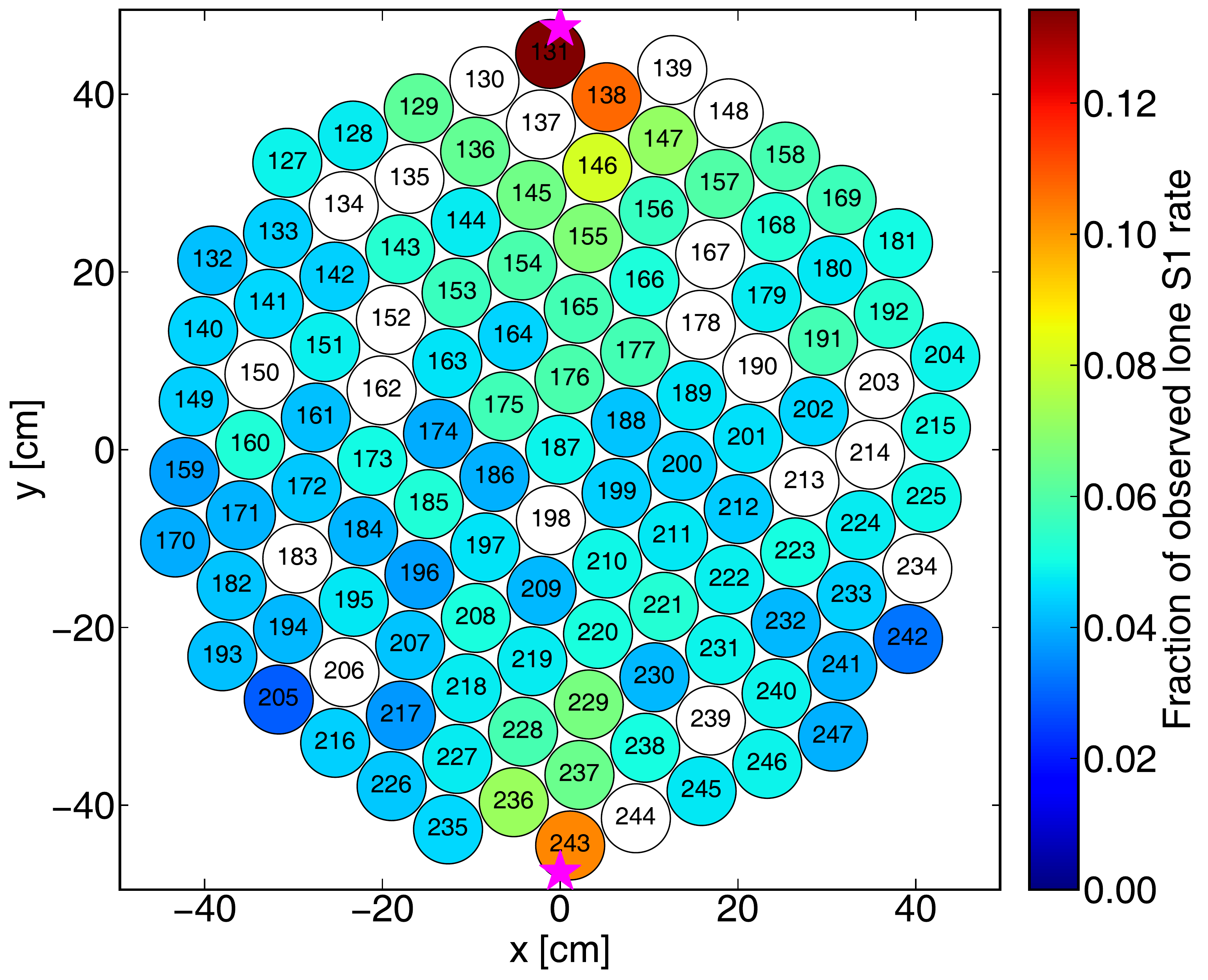}
    \caption{Fraction of lone S1 signals (color scale) between 0 and $70\,$PE observed by each PMT (circles) in the bottom array during $^{220}$Rn calibrations. Numbers in the circles indicate PMT IDs. Magenta stars mark the points where internal calibration sources are injected into the detector below the cathode. White circles represent PMTs that were non-functional in SR1.}
    \label{fig:cuts:InjectionFraction}
\end{figure}

The first publication of XENON1T results \cite{xe1t_first} featured a component of the background model that was flat in cS1 vs. cS2$_\mathrm{b}$ space. It was motivated by the observation of events in $^{220}$Rn calibration data that featured cS2$_\mathrm{b}$ values smaller than expected for regular ERs \cite{xe1t_first}. Those events are referred to as anomalous leakage. The increased statistics in calibration data acquired in SR1 enabled more detailed studies and a better understanding of anomalous leakage events. Selection criteria were developed to remove this type of events from $^{220}$Rn calibration data, obviating the inclusion of an anomalous component to the background models~\cite{xe1t_comb}:
\begin{itemize}
    \item $^{220}$Rn calibration data has an increased lone S1 rate in two regions below the cathode where the isotope is injected into the TPC. These S1 signals can get accidentally paired to lone S2 signals. 
Fig.\,\ref{fig:cuts:InjectionFraction} shows the fraction of lone S1 signals between 0 and $70\,$PE observed by each PMT in the bottom array during SR1 $^{220}$Rn calibrations. Events for which the fraction of the S1 signal observed by the PMTs (131, 138, 146, 147, 243, 236 and 237) close to the injection points (magenta star) exceeds an S1 signal size-dependent threshold are rejected. The threshold is defined such that 99\% of $^{220}$Rn events that pass all previously described selection criteria are accepted.
\item Additionally, lone S1 signals feature larger widths in time if they originate from mis-identified single electron signals. The width where 90\% of the signal peak is contained is required to be smaller than an S1 signal size-dependent threshold. The threshold is optimized in the same way as the criterion for the $^{220}$Rn injection points.
\end{itemize}
The $^{220}$Rn-specific selection criteria have been conservatively applied to background data in order to exclude potential artefacts remaining in datasets acquired shortly after $^{220}$Rn calibration periods due to the remaining isotopes inside the connection pipes. The combined acceptance of physical interactions is $(98\pm1)$\%.

\subsection{Signal Acceptances}

\begin{figure*}
    \centering
    \includegraphics[width = 0.95\textwidth]{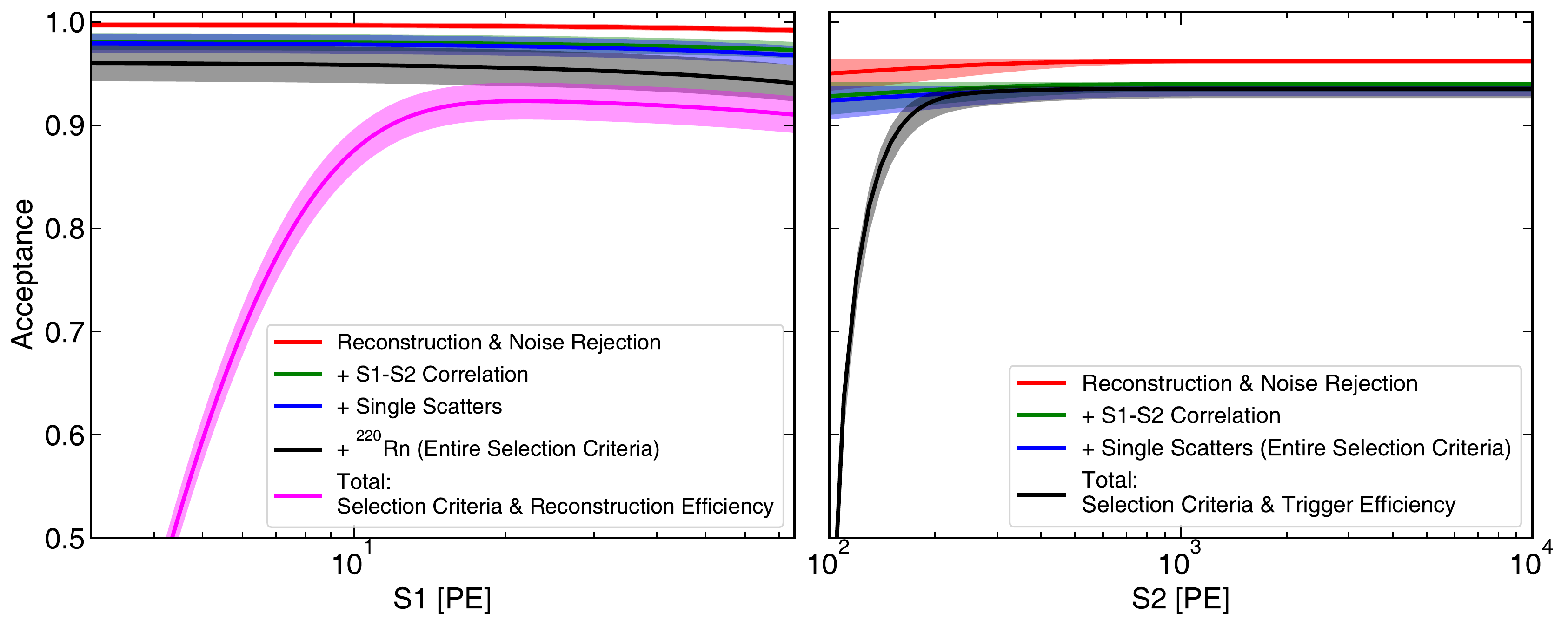}
    \caption{Evolution of the acceptance of S1 and S2 signals from physical interactions by incremental application of the three categories of selection criteria described in the text as a function of uncorrected S1 and S2 signal sizes. The total acceptance is also shown after considering the reconstruction efficiencies introduced in Sec.\,\ref{sec:paxfax}.}
    \label{fig:cuts:acceptances}
\end{figure*}


Fig.\,\ref{fig:cuts:acceptances} shows the cumulative signal acceptances of the described categories of selection criteria as functions of uncorrected S1 (left) and S2 (right) signal sizes. The smooth curves are determined by fitting a first order polynomial function to the data points in the S1 signal space and a function of the form $(a+b\cdot \mathrm{S2})(1-c\cdot\mathrm{exp}(-\mathrm{S2}/d))$ in the S2 signal space. The uncertainty bands account for the statistics present during the acceptance estimation from data as well as for the systematics of the fit which are derived from the deviation of the data points from the best fit line. Correlations between most of the selection criteria are found to be negligible and the acceptances are estimated from control samples in background and calibration data. Four requirements on S1 signals (signal size contributed by a single PMT, likelihood of PMT hit pattern given the reconstructed position, consistency of signal fraction seen in the top PMT array with the interaction depth and the single scatter criteria) as well as the $^{220}$Rn specific selection criteria show correlations with other criteria. Hence, their acceptances are conservatively estimated by applying them consecutively to $^{220}$Rn calibration data after all other criteria.


In the S1 signal space, the S1-S2 signal correlation conditions and the $^{220}$Rn specific criteria have the largest impact and the entire selection acceptance is given by the black curve. In the S2 signal space, the acceptance is primarily influenced by the data quality and the S1-S2 signal correlation criteria. The $^{220}$Rn specific criteria have no impact since they do not affect S2 signals. The entire signal selection acceptance averaged over the region of interest yields $(91\pm4)$\%.

The total acceptance were overlaid together in Fig.\,\ref{fig:cuts:acceptances} after considering the reconstruction efficiencies that introduced in Sec\,\ref{sec:paxfax}, which were passed to the statistical inference for dark matter searches~\cite{xe1t_long_analysis_2}.

%% file: Discrimination.tex
\section{Optimization of the fiducial volume}
\label{sec:cuts:fiducial}

\begin{figure*}
    \centering
    \includegraphics[width = 0.45\textwidth, height = 6cm]{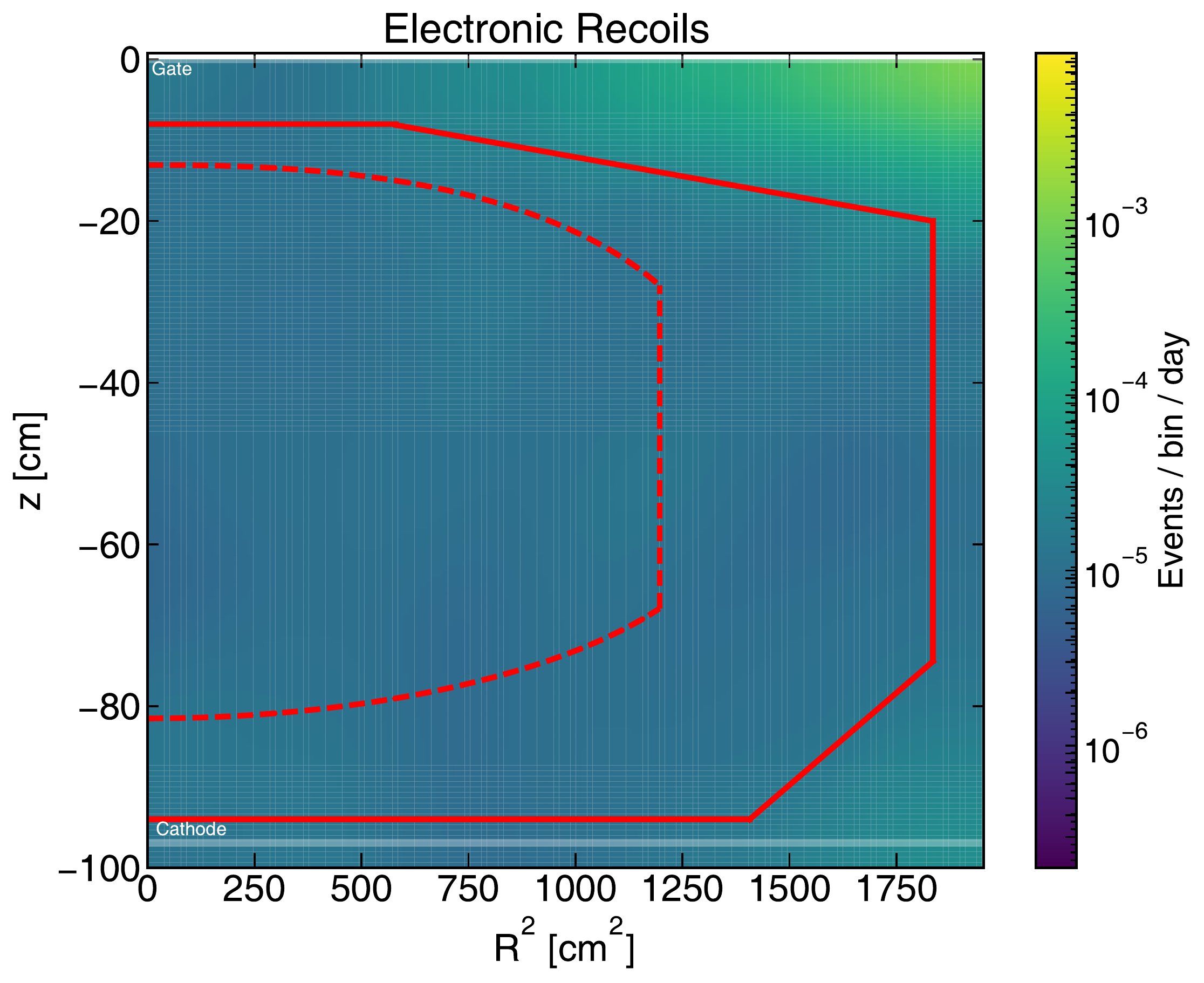}
    \includegraphics[width = 0.45\textwidth, height = 6cm]{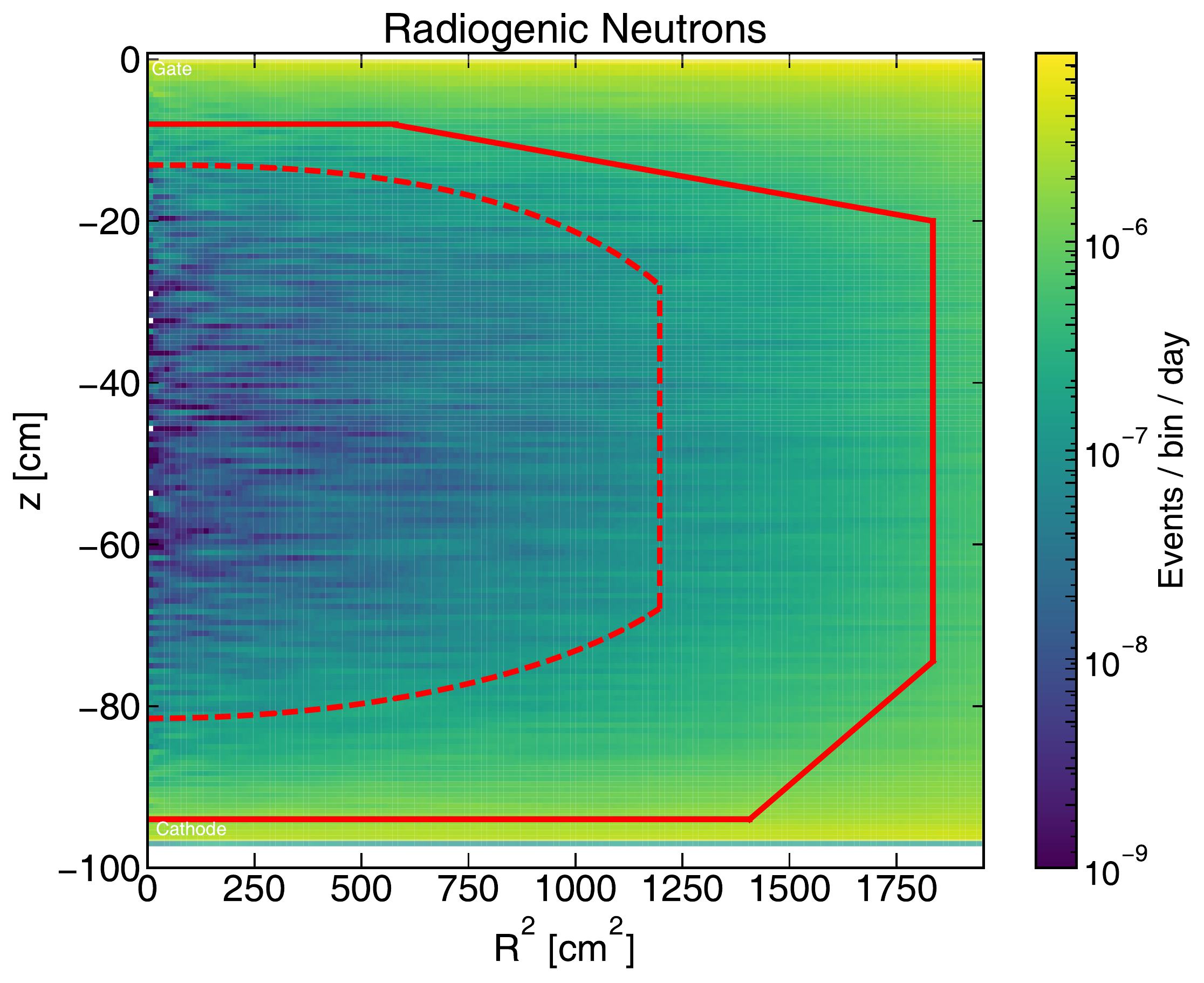}\vspace{0.2cm}
    \includegraphics[width = 0.45\textwidth, height = 6cm]{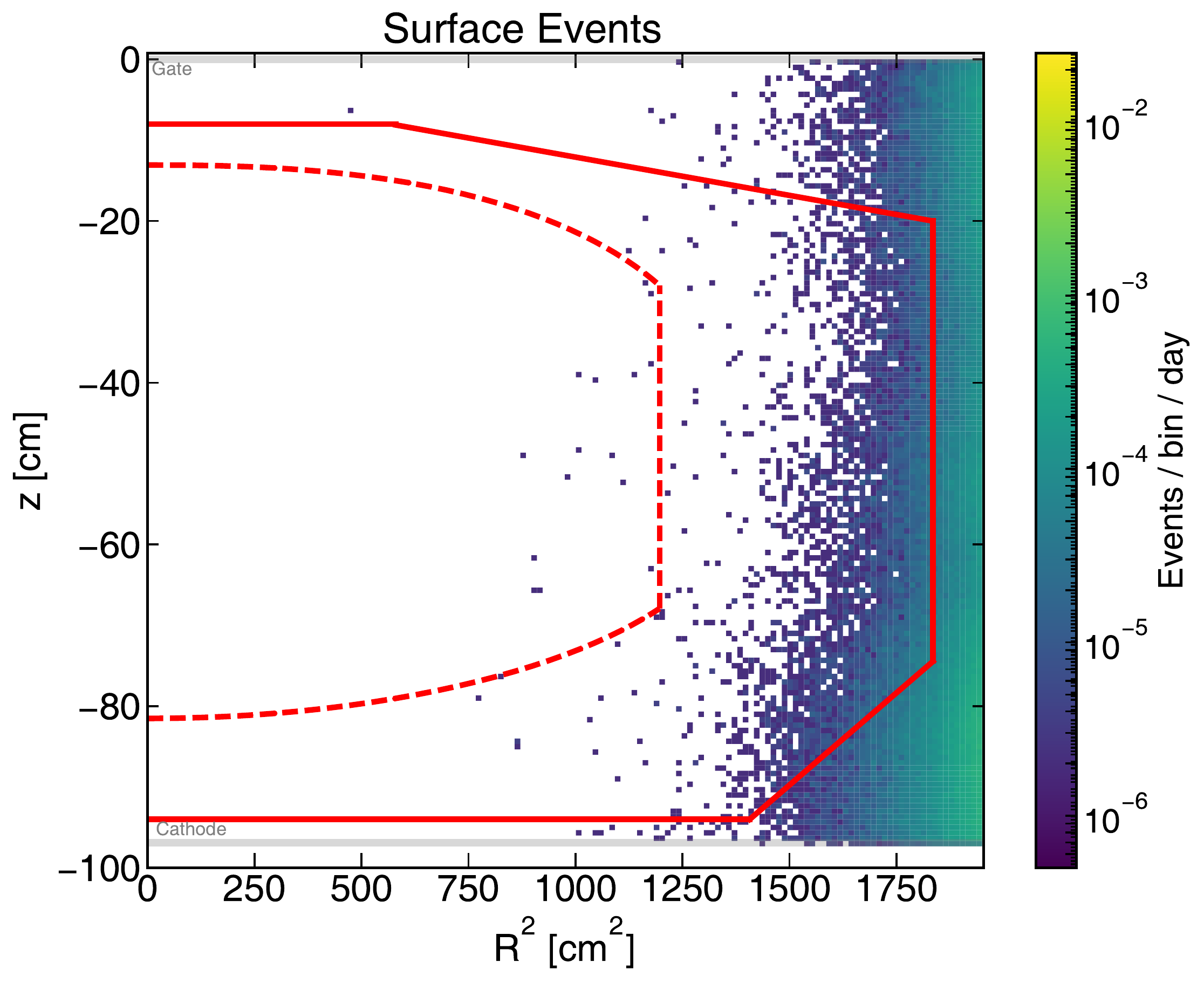}
    \includegraphics[width = 0.45\textwidth, height = 6cm]{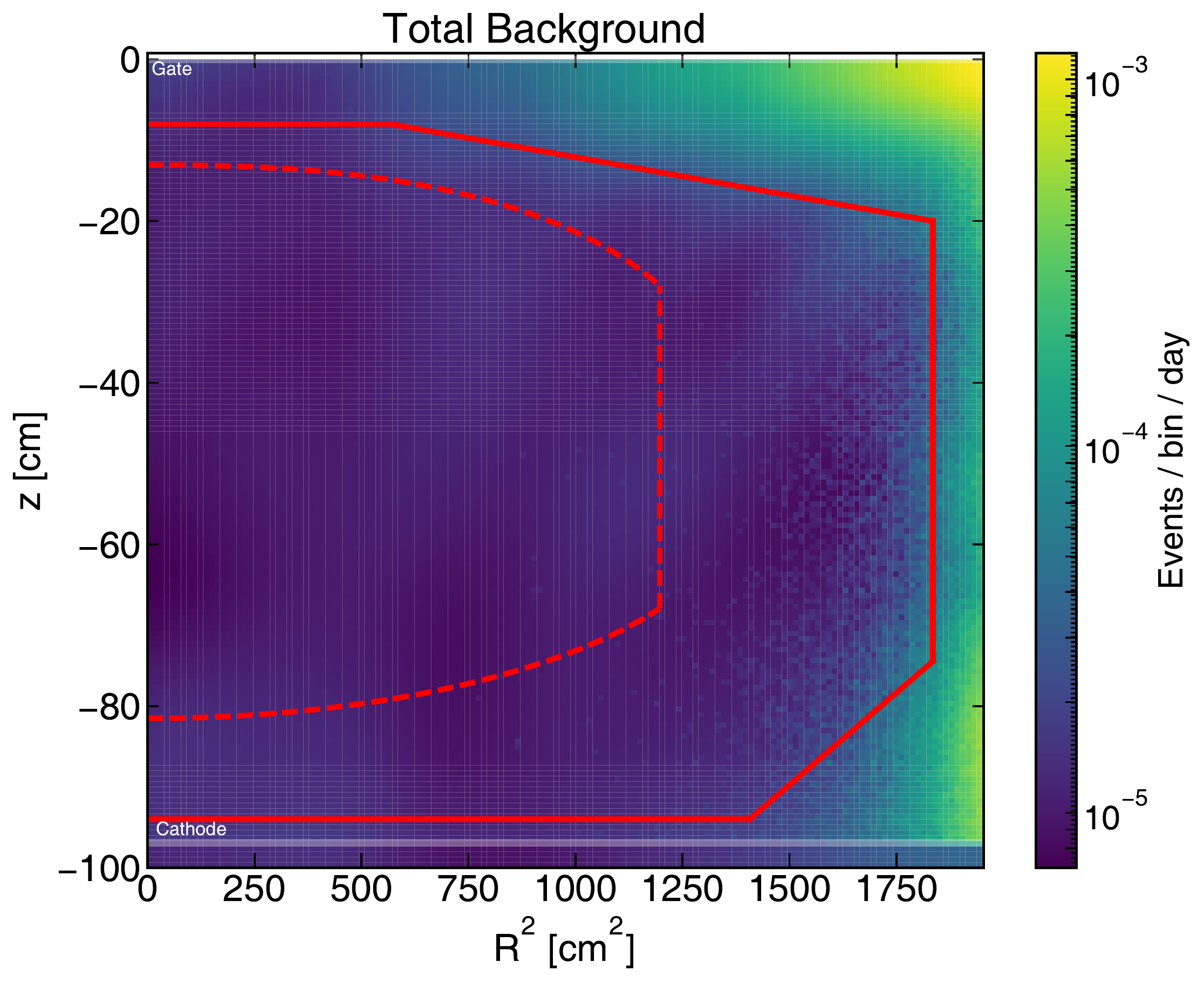}
    \caption{Spatial distribution of modeled background rates (color scale) from ER (top left), radiogenic neutrons (top right), surface events (bottom left) and total background (bottom right) considering the cS1-cS2$_\mathrm{b}$ region in which a $50\,$GeV WIMP signal is expected to have the highest significance compared to the total background. Note that the rate axis is not the same for all panels. Red solid lines mark the result of the FV optimization ($1.3\,$\ton) and red dashed lines the inner core mass ($0.65\,$\ton).}
    \label{fig:cuts:fv}
\end{figure*}

The fiducial volume (FV) is optimized using the $[R, z]$ distributions of the background components, shown in Fig.\,\ref{fig:cuts:fv}. The distributions are evaluated considering only the cS1-cS2$_\mathrm{b}$ region in which a $50\,$GeV/$c{^2}$ WIMP signal is expected to have the highest significance compared to the total background. Additionally the models are limited to $R<44\,$cm in order to exclude biases from signal corrections related to the decreasing position resolution beyond that region (Sec.~\ref{pos_res}).
ER background (top left) is mostly from homogeneously distributed $^{214}$Pb $\beta$-decays and inhomogeneously distributed emission from detector materials and is modeled with kernel density estimation (TKDE class from the ROOT framework~\cite{ROOT}) using blinded WIMP search data. Charge losses exhibited by surface events (bottom left) produce artificially small S2 signals, which can be mis-reconstructed further inside the TPC. These events are modeled using side-bands just outside the DM search region. NR events (top right) are only expected from neutron emission of radio-impurities in the detector materials and are modeled with a MC toolkit based on GEANT4~\cite{geant4}. The simulated distributions are scaled to the expected number of background events predicted with the SOURCES-4A simulation package \cite{sources-4a, sources-4a-modification} and verified in data with multiple-scatter NR events. The bottom right panel in \figref \ref{fig:cuts:fv} shows the total background distribution with all components scaled to their expected relative intensities.

The FV is restricted in depth to $z\in[-94,-8]\,$cm in order to exclude mis-reconstructed events from the gas volume and events originating near the cathode, where the electric field exhibits a higher non-uniformity. The maximum radius is set as $42.84\,$cm in order to exclude biases from field inhomogeneity and to reject the bulk of surface background events. The radial boundary could have been moved slightly inward to reject almost all surface background events, but it was intentionally set such that about 100 surface events are included in order to provide enough statistics to scale the background model~\cite{xe1t_long_analysis_2}. Because the only spatial variable used in statistical inference is $R$, the outer FV regions are further restricted in $z$ such that each bin in $R$ exhibits a background rate homogeneous within 10\%. The result of the optimization is represented by the red solid lines in Fig.\,\ref{fig:cuts:fv}.

As shown in Fig.\,\ref{fig:cuts:fv} top right, the NR background distribution has a strong $z$ dependence at inner radii, which motivated further segmentation of the volume and defining a core volume (dashed line). This region was constructed by maximizing the WIMP signal over square root of background and has a $\sim$80\% lower neutron rate than the total FV. 

The fiducial mass contained in the total FV is determined by two methods: from the ratio of $^{83m}$Kr calibration events in the selected volume with respect to the total sensitive volume and from geometrical calculation. The relative difference between the two methods increases from 1.6\% to 1.9\% with increasing radius due to increasing position reconstruction uncertainties. The two values are combined and their difference is taken into account by a systematic uncertainty. The FV contains $(1.30\pm0.01)\,$t of liquid xenon at -92 $^\circ$C while the core mass makes up half of the fiducial mass. 

\section{Signal and Background Discrimination}
\label{sec:Discrimination}

\begin{figure*}
    \centering
    \includegraphics[width = 0.7\textwidth]{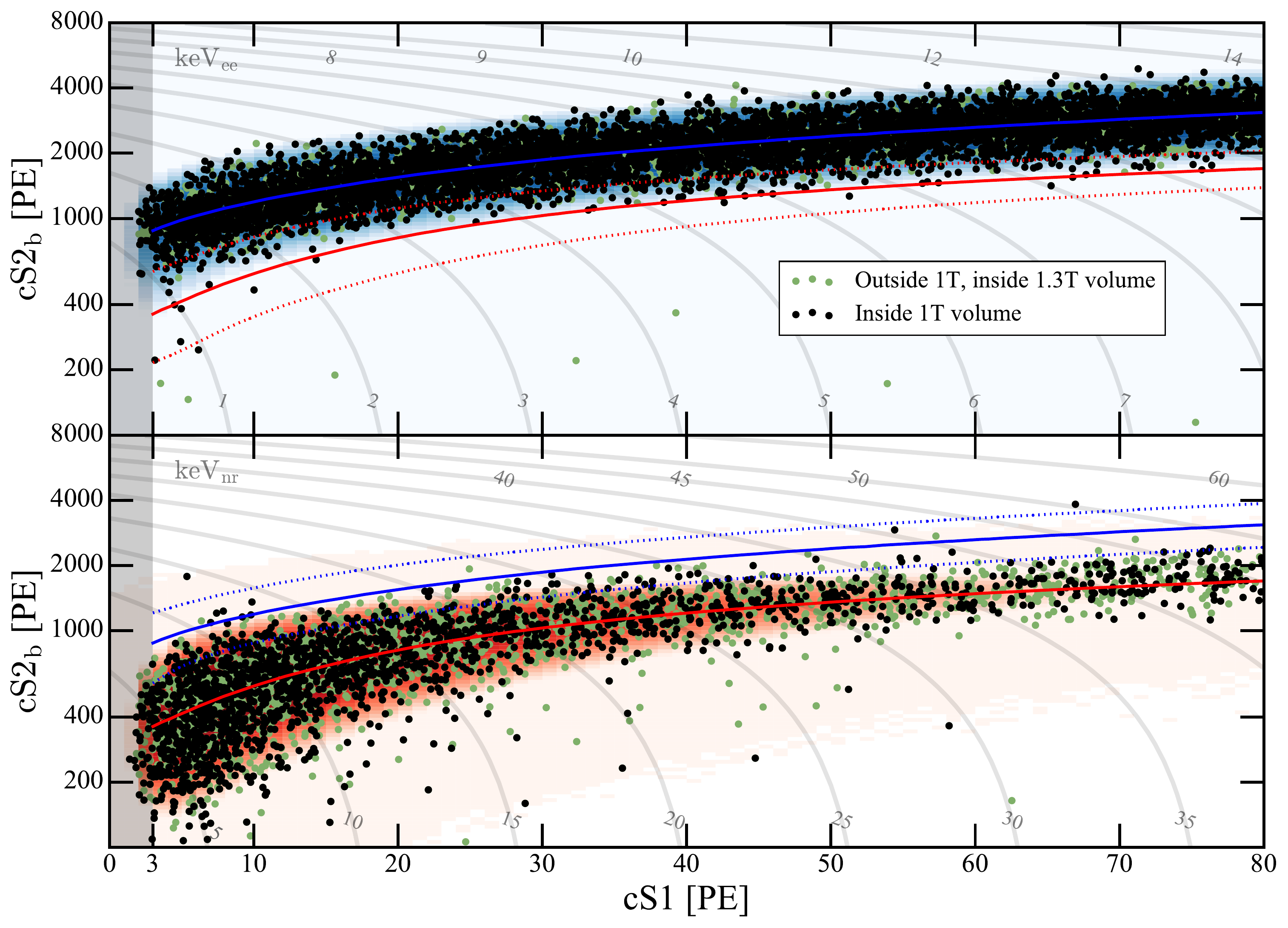}
    \caption{$^{220}$Rn (upper panel) and neutron generator (lower panel) calibration data in ER/NR discrimination space for events within the inner $1~\,$t cylindrical volume (black) and outside the $1~\,$t volume but within the $1.3~\,$t  volume (green). The models extracted from the calibration data for ER (blue) and a $200\,$GeV/$c{^2}$ WIMP (red) are shaded and additionally drawn as 10\%-50\%-90\% (dotted-solid-dotted) contour lines. The energy axes for ER (top) and NR (bottom) are drawn as grey contours. More details on background and signal modeling are given in \cite{xe1t_long_analysis_2}.}
    \label{fig:Discrimination}
\end{figure*}

The detector response to ERs is modeled with a Markov Chain Monte Carlo (MCMC) fit to $\beta$-decays of $^{212}$Pb selected from $^{220}$Rn calibration data \cite{xe1t_long_analysis_2}. The fit uses 1867 (14138) events
in SR0 (SR1) reconstructed at radii smaller than $34.6\,$cm in order to reject all external and surface backgrounds from the training sample that feature a loss in the charge signal. This smaller volume contains a fiducial mass of $\sim1\,$t.
The same method is applied to model the response to NRs with 3247 (4889) events in SR0 (SR1) from $^{241}$AmBe and neutron generator calibration data.

Figure.\,\ref{fig:Discrimination} shows $^{220}$Rn (top) and neutron generator (bottom) calibration data that form band-shaped distributions in the cS2$_\mathrm{b}$ vs. cS1 WIMP search region of interest. The extracted models are indicated by shaded colored regions together with the respective 10\%-50\%-90\% (dotted-solid-dotted) contour lines. Colored points differentiate events found in the $1\,$t FV (black) or outside the $1\,$t FV but inside the $1.3\,$t FV (green). Surface events visible below the ER band are primarily located in the outer detector region. Because the neutron generator is located outside the TPC, the NR calibration data set features a larger fraction of events at larger radii.

The ability to discriminate between ER and NR processes is crucial for background reduction and therefore for the experimental sensitivity to WIMP dark matter. The separation of the ER and NR bands is quantified by the expected fraction of ER events reconstructed below the median of the NR distribution (50\% NR acceptance, solid red line in Fig.\,\ref{fig:Discrimination}). According to the extracted models the leakage fraction yields $(0.3\pm0.1)$\% 
between 3 and 70 PE in cS1 and within the $1.3\,$t FV.
In addition to ER background originating from $^{214}$Pb $\beta$-decays homogeneously distributed within the detector, radon progeny accumulated on the PTFE surface and radiogenic backgrounds from materials can also contaminate the signal region, making the radial position another strong discriminator between signal and background (Sec.\,\ref{sec:cuts:fiducial})~\cite{xe1t_long_analysis_2} .




%% file: Outlook.tex
\section{Summary and Outlook}
\label{sec:Outlook}

In this article we described details on signal reconstruction, event selection and calibrations used to search for elastic WIMP-nucleon interactions in XENON1T \cite{xe1t_comb,xe1t_SD,xe1t_pion}. 
Most methods shown here are also being used in current and future results for WIMP and alternative dark matter models as well as other low background searches.\\
The experiment was operated under stable conditions for more than one year. Periodic calibrations with internal sources such as $^{83\mathrm{m}}$Kr and $^{220}$Rn 
are used to model time-dependent parameters such as electron live time and progressing electric field distortion due to charge accumulation on PTFE surfaces.

Compared to the predecessor experiment XENON100~\cite{xe100_analysis}, the MC simulation efforts were significantly strengthened in the analysis chain of XENON1T. Simulated waveforms 
are used to compute the peak finding efficiencies, peak reconstruction biases, and event reconstruction performance of the data processor. MC simulations for light propagation within the TPC are employed to tune position reconstruction algorithms and to determine a goodness of fit between expected and measured patterns on the PMT arrays.

XENON1T was decommissioned in December 2018
and the imminent detector upgrade XENONnT is being constructed. XENONnT will feature an increased target mass of $5.9\,$\ton and a sensitivity enhanced by more than an order of magnitude \cite{xe1t_mc}. First steps towards this goal were realized by upgrades of the XENON1T purification system after SR1. The improvements 
resulted in the highest electron lifetime achieved in the experiment of $\sim1\,$ms and a reduction of the dominating background from $^{222}$Rn of about 50\%. 


